\documentclass{elsart}

\usepackage{amsmath}
\usepackage{amssymb}
\usepackage{graphicx}

\usepackage{natbib}

\newcommand{\Ub}{U_\text{b}}
\newcommand{\kav}{k_\text{av}}
\newcommand{\wav}{w_\text{av}}
\newcommand{\mub}{\mu_\text{b}}

\newcommand{\Vark}{\text{Var}[k]}
\newcommand{\fkz}{f_{k_0}}

\begin{document}

\begin{frontmatter}

\title{The traveling wave approach to asexual evolution: Muller's ratchet and 
speed of adaptation}

\author[first,cor]{Igor M. Rouzine}
\author[second]{\'Eric Brunet}
\author[third]{Claus O. Wilke}

\address[first]{Department of Molecular Biology and Microbiology,\\
Tufts University, 136 Harrison Avenue, Boston, MA 02111}
\address[second]{Laboratoire de Physique Statistique,\\
\'Ecole Normale Sup\'erieure, 24 rue Lhomond,\\
75230 Paris C\'edex 05, France}
\address[third]{Section of Integrative Biology,\\
Center for Computational Biology and Bioinformatics,\\
and Institute for Cell and Molecular Biology,\\
University of Texas, Austin, TX 78712, USA}
\corauth[cor]{Corresponding author. email: igor.rouzine@tufts.edu. phone: 617-636-
6759. fax: 617-
636-4086}

\begin{abstract}
We use traveling-wave theory to derive expressions for the rate of accumulation of 
deleterious mutations under Muller's ratchet and the speed of adaptation under 
positive selection in asexual populations. Traveling-wave theory is a
semi-deterministic description of an evolving population, where the bulk of the 
population is modeled using deterministic equations, but the class of
the highest-fitness genotypes, whose evolution over time determines loss or
gain of fitness in the population, is given proper stochastic treatment. We derive improved
methods to model the highest-fitness class (the stochastic edge) for both Muller's ratchet 
and adaptive evolution, and calculate analytic correction terms that compensate 
for inaccuracies which arise when treating discrete fitness classes as a 
continuum. We show that traveling wave theory makes excellent predictions for the 
rate of mutation accumulation in the case of Muller's ratchet, and makes good 
predictions for the speed of adaptation in a very broad parameter range. We 
predict the adaptation rate to grow logarithmically in the population size until 
the population size is extremely large.
\end{abstract}

\end{frontmatter}

\section{Introduction}

\begin{quote}
``I was observing the motion of a boat which was rapidly drawn along a narrow 
channel by a pair of horses, when the boat suddenly stopped---not so the mass of 
water in the channel which it had put in motion; it accumulated round the prow of 
the vessel in a state of violent agitation, then suddenly leaving it behind, 
rolled forward with great velocity, assuming the form of a large solitary 
elevation, a rounded, smooth and well-defined heap of water, which continued its 
course along the channel apparently without change of form or diminution of 
speed.'' \citep{Russel1845}
\end{quote}

One of the fundamental models of population genetics is that of a finite, 
asexually reproducing population of genomes consisting of a large number of sites 
with multiplicative contribution to the total fitness of the genome. This model 
has been studied for decades, and has presented substantial challenges to 
researchers trying to solve it analytically. Even in its most basic formulation, 
where each site is under exactly the same selective pressure, the model has not 
been fully solved to this day. For the special case of vanishing back mutations, 
the model reduces to the problem of Muller's ratchet 
\citep{Muller64,Felsenstein74}. A tremendous amount of research effort has been 
directed at this problem 
\citep{Haigh78,Pamiloetal87,Stephanetal93,HiggsWoodcock95,GordoCharlesworth2000a,
GordoCharlesworth2000b,Rouzineetal2003}. Other special cases of this model are 
mutation--selection balance when the forward and back mutation rates are equal 
\citep{WoodcockHiggs96}, and the speed of adaptation under various conditions 
\citep{Tsimringetal96,Kessleretal97,GerrishLenski98,Orr2000,Rouzineetal2003,
Wilke2004}.

In 1996, \citeauthor{Tsimringetal96} pioneered a new approach to studying the 
multiplicative multi-site model. They described the evolving population as a 
localized traveling wave in fitness space, using partial differential equations 
developed to describe wave-like phenomena in physical systems. A traveling wave is 
a localized profile traveling at near-constant speed and shape (physicists refer 
to such phenomena also as \emph{solitary waves}). We can envision a population as 
a traveling wave of the distribution of the mutation number over genomes if the 
relative mutant frequencies in the population stay approximately constant while 
the population shifts as a whole. For example, a population may have specific 
abundances of sequences at one, two, three, or more mutations away from the least 
loaded class at all times, but the least-loaded class moves at constant speed by 
one mutation every ten generations.
 
Encouraging results by \citet{Tsimringetal96} were based on two strong 
approximations. Firstly, all fitness classes, including the best-fit class, were 
described deterministically, neglecting random effects due to finite population 
size. Finite population size was introduced into the problem as a cutoff of the 
effect of selection at the high-fitness edge, when the size of a class becomes 
less than one copy of a genome. Secondly, \cite{Tsimringetal96} approximated the 
traveling wave profile with a function continuous in fitness (or mutation load). 
In these approximations, \citet{Tsimringetal96} demonstrated the existence of a 
continuous set of waves with different speeds. The cutoff condition determined the 
choice of a specific solution and the dependence of the speed on the population 
size.

\citet{Rouzineetal2003} confirmed the qualitative conclusions by 
\citet{Tsimringetal96} and refined their quantitative results in two ways, by 
taking into account the random effects acting on the smallest, best-fit class, and 
showing that, in a broad parameter range, approximating the logarithm
of the wave profile as a smooth function of the fitness is a much better
approximation than approximating the wave profile itself as a smooth
function. It was shown that the substitution rate increases 
logarithmically with the population size, until the deterministic single-site limit 
is reached at extremely large population sizes and the theory breaks down.

The purpose of the present paper is to show that the results by 
\citet{Rouzineetal2003} can still be improved regarding both the treatment of the 
high-fitness edge and the deterministic part of the fitness distribution. Because 
the original work was presented in an extremely condensed format, we will here re-derive the
general theory in detail. Then, we will present improved treatments of 
the stochastic edge that lead to accurate predictions for the substitution rate defined as the
average gain in beneficial alleles per genome per generation. We consider in detail
two opposite parametric limits, Muller's ratchet (when beneficial mutation events 
are not important) and adaptive evolution (when deleterious mutations are not 
important). 

Because the range of validity of our approach caused some confusion in the literature, we discuss it in detail in the main text and Appendix. Briefly, both in Muller's ratchet and in the adaptation
regime, we assume that the total number of sites is large, and the selection coefficient $s$ is
small. The population size $N$ should be sufficiently large, so that  the difference in the
mutational load between the least-loaded and average genomes is much larger than 1. In other
words, a large number of sites are polymorphic at any time. For the adaptive evolution, the condition corresponds to the average substitution rate $V$ being much larger than  $s/\ln(V/\Ub)$, where $\Ub$ is the beneficial mutation rate, or population sizes being much larger than $\sqrt{s/U_b^3}/\ln(V/U_b)$.
We also assume that  $V$ is much larger than $U_b$. In smaller populations, adaptation occurs by
isolated selective sweeps at different sites, and one-site models apply.  Two-site models of
adaptation, such as the clonal interference theory \citep{GerrishLenski98,Orr2000, Wilke2004}, can
be used to describe the narrow transitional interval in $N$.  Further, if the deleterious mutation
rate per genome is much larger than $s$, and the average population fitness is sufficiently high, an
additional broad interval of $N$ appears, where deleterious mutations accumulate (Muller's ratchet).
Using numeric simulations and analytic estimates, we demonstrate good accuracy of our results in a
very broad parameter range relevant for various asexual organisms, including asexual RNA and DNA
viruses, yeast, some plants, and fish.

The manuscript is organized as follows. In Section 2, we describe the model and the general method to derive evolutionary dynamics in terms of the fitness distribution. In Section 3 and 4, we consider in detail two particular cases, Muller's ratchet and the process of adaptation, respectively, and test analytic results with computer simulations. In Section 4, we discuss our findings.

\section{Traveling-wave theory}

\subsection{Model assumptions}

We consider a multi-site model of $L$ sites, where each site can be in two states,  
i.e., carry one of two alleles, either advantageous or deleterious. The 
deleterious allele reduces the overall fitness of a genome by a factor of $1-s$, 
where $s\ll 1$ is the selective disadvantage per site. We assume that there are no 
biological interactions between sites (epistasis), so that the fitness of a genome 
with $k$ deleterious alleles (mutational load) is $(1-s)^k\approx e^{-ks}$. We 
refer to the frequency of sequences with mutational load $k$ in the population as 
$f_k$, and write the population-average fitness as $\wav=\sum_k e^{-sk} f_k$. We 
introduce $\kav$, which is mutational load for which a sequence's fitness is 
exactly equal to the population mean fitness, as given by $\wav=e^{-s\kav}$. For 
small $s$, $\kav$ is approximately the average mutational load in the population, 
$\kav\approx \sum_k k f_k$. We denote the mutational load of the best-fit sequence 
in the population as $k_0$. Note that in general $k_0\neq 0$, that is, the best-
fit sequence in the population is not the sequence with the overall highest 
possible fitness.

We assume that an allele can mutate into an opposite allele with a small 
probability $\mu$. For the sake of simplicity of the derivation, we assume that the 
mutation rate is low, so that there is, at most, one mutation per genome per 
generation. The case when multiple mutations per genome are frequent takes place 
at very large population sizes when the average substitution rate is high, larger 
than one new beneficial allele per generation. \citet{Rouzineetal2003} considered 
this more general case and showed that there is no essential change in the final 
expression for the substitution rate. Thus, for finite population size, the 
assumption of a single mutation per genome per round of replication is not limiting (see also the next subsection).

In real genomes, $s$ varies between sites. Moreover, 
\citet{Gillespie83,Gillespie91} and \citet{Orr2003} argued that the distribution 
of $s$ differs between sites with beneficial and deleterious alleles. A viable 
genome represents a highly-fit, non-random selection of alleles, so that 
deleterious mutations should generally have larger effects than beneficial 
mutations. For the same reason, these authors predicted that the effective 
distribution of $s$ for beneficial alleles should have a universal exponential 
form. In the present work, we do not consider variation in $s$. Instead, we use a simplified model including only those sites into the 
total number of sites $L$ ---with either deleterious and beneficial alleles---
whose selection coefficient is on the order of the same typical value $s$, and 
approximate all selection coefficients at these sites with a constant $s$. The 
choice of $s$ and, hence, of the set of included sites depends on the time scale 
of evolution under consideration. Strongly deleterious mutations with effects much 
larger than $s$ are cleared rapidly from a population. Strongly beneficial 
mutations are fixed at the early stages of evolution. Note that in the ``clonal 
interference'' approximation \citep{GerrishLenski98}, which considers competition 
between two beneficial clones emerging at two sites, variation of $s$ must be 
taken into account to make continuous adaptation possible. In contrast, in the 
present theory, which allows new beneficial clones to grow inside of already 
existing clones, the importance of variation in $s$ is less obvious. We hope to 
address this matter elsewhere.

We discuss the validity of our approach in detail for the limits of Muller's ratchet and adaptation 
and give a summary of the central simplifications---numbered 1 through 6 and 
referenced throughout the text---in the Appendix. Note that we can verify the 
validity of the various assumptions only after the fact, once we have obtained our 
final results. All simplifications are asymptotically exact, i.e., are based on the existence of small dimensionless parameters. The most limiting requirements are that the high-fitness tail of the distribution is long, and that the distribution is far from the unloaded and fully loaded (possible best-fit and less-fit) genomes. 

\subsection{General approach}

The first idea underlying the approach of "solitary wave" is to classify all 
genomes in a population according to their fitness (mutational load), regardless 
of specific locations of deleterious alleles in a genome, and focus on evolution 
of fitness classes. The second idea is to describe evolution of most fitness 
classes deterministically. To take into account the effects of finite population 
size, such as genetic drift and randomness of mutation events, only one class with 
the highest fitness is described stochastically using the standard two-allele 
diffusion approach. The best-fit class is considered a minority "allele" in a 
population, and all other sequences are considered the majority "allele". The 
reason why stochastic effects can be neglected already for the next-to-best class 
is that, in a broad parameter range, the fitness distribution decreases 
exponentially towards the stochastic edge. Hence, the next-to-best class is large 
enough to neglect stochastic effects, especially in the adaptation regime (see 
estimates in Sections 3 and 4).  

We note that neglecting stochastic effects completely and considering 
the limit of infinite population is not correct. 
As we show below, stochastic processes 
acting on the best-fit class limit the overall evolution 
rate and make it dependent on the population size. Even for a modest number of 
sites ($L$ = 15-20), the substitution rate is predicted to reach the true 
deterministic limit only in astronomically large populations not found in nature 
\citep{Tsimringetal96,Rouzineetal2003,DesaiFisher2007}. For the same reason, the 
assumption of one mutation per genome we made in our model is not a limiting 
factor and, as shown by \citet{Rouzineetal2003}, does not change much in the final expression for the average substitution rate.

The formal procedure consists of several steps, as follows.

 (i) The frequencies of all fitness classes excluding the best-fit class are described by a deterministic balance equation. 
 
 (ii) The  equation is shown to have a traveling wave solution with an arbitrary speed (the average substitution rate). 
 
 (iii) The leading front of the wave (high-fitness tail of the distribution) is shown to end abruptly at a point, expressed in terms of the wave speed.
 
 (iv) The difference between the values of the fitness distribution at the center and the edge is expressed in terms of the wave speed.
 
 (v) The value at the center is found from the normalization condition. 
 
 (vi) Because the biological justification for the lack of genomes beyond the high-fitness cutoff is finite size of population, the cutoff point is identified with the stochastic edge.
  
(vii) To determine the wave speed as a function of the population size, the average frequency of the least-loaded class is estimated from the classical diffusion result and matched to the deterministic cutoff value.

\subsection{Equation for the deterministic part of the fitness distribution}

We proceed with the first step. On the basis of our model assumptions and neglecting 
multiple mutation events per generation per genome, we can write the deterministic 
time evolution of the frequency $f_k(t)$ of genomes with mutational load $k$ as
\begin{align}\label{eq:det-one-mutation}
	f_k(t+1) &= \frac{1}{\wav(t)}\Big(e^{-s(k-1)}\mu(L-k+1)f_{k-1}(t)\notag\\
		&\quad +e^{-s(k+1)}\mu(k+1)f_{k+1}(t)
		 + e^{-sk}(1-\mu L)f_k(t)\Big)\,,
\end{align}
where $k$ runs from 0 to $L$ and $f_{-1}(t)\equiv f_{L+1}(t) \equiv 0$. By 
definition, $\sum_k f_k(t)=1$ for all times $t$. Now we introduce the total per-genome mutation rate $U = \mu{}L $, and the ratio of beneficial mutation rate per 
genome to the total mutation rate, $\alpha_k=\mu{}k/U=k/L$. Inserting these 
expressions into Eq.~\eqref{eq:det-one-mutation}, expressing $\wav$ in terms of 
$\kav$, expanding $e^{-sx}\approx 1-sx$ (which we are allowed to do under the 
condition that $s|k-\kav|\ll 1$, Simplification~1), and neglecting all terms 
proportional to $sU$, we arrive at:
\begin{align}
  f_k(t+1) &= U(1-\alpha_{k-1})f_{k-1}(t) + U\alpha_{k+1} f_{k+1}(t)\notag\\
	&\qquad\qquad\qquad\qquad +[1-U-s(k-\kav)]f_k(t)\,.
\end{align}
As mentioned before, we consider the case when the traveling wave is far from the 
sequence with the highest possible fitness, $k=0$, as given by the condition $|k-
k_{\rm av}| \ll k_{\rm av}$. Therefore, $\alpha_k$ depends only slowly (linearly) 
on $k$, we are allowed to replace $\alpha_k$ by $\alpha\equiv \alpha_{\kav}$ 
(Simplification~2), and find
\begin{equation}\label{eq:det-expanded-Ualpha}
 f_k(t+1) - f_k(t) = U(1-\alpha)f_{k-1}(t)+U\alpha f_{k+1}(t)-[U+s(k-
k_\text{av})]f_k(t)\,.
\end{equation}

Our goal is to turn this expression into a continuous differential equation. As 
the mutation rates are low, $f_k(t)$ evolves very slowly in time, and we can write 
$f_k(t+1) \approx f_k(t) + \partial f_k(t)/\partial t$ (Simplification~3). 

We need 
to be more careful when making a continuous approximation for $f_{k}(t)$ as a 
function of $k$. As we show below, $f_k(t)$ changes rapidly with $k$ in the 
important high fitness tail. Therefore, the Taylor expansion of $f_k(t)$ is not 
justified. However, the logarithm of $f_k(t)$ is a smooth function of $k$ in a 
broad parameter range, provided the "lead" of the distribution is large (Simplification~4).
[The lead is the difference in number of mutations from the population center to the least-loaded class in the population \citep{DesaiFisher2007}.]
Therefore, a better approximation, which 
represents an improvement on the work by \citet{Tsimringetal96}, is to do the 
Taylor expansion on $\ln f_k$ and write $f_{k+1}(t)=f_k(t) \exp[\partial \ln 
f_k(t)/\partial k]$. With these approximations, and after introducing a rescaled 
time $d\tau=Udt$ and a rescaled selection coefficient $\sigma=s/U$, we find
\begin{equation}\label{eq:final-wave-eq}
  \partial \ln f_k(t)/\partial \tau = (1-\alpha) e^{-\partial\ln f_k(t)/\partial 
k}
	+ \alpha e^{\partial\ln f_k(t)/\partial k}-\sigma(k-\kav)-1\,.
\end{equation}
This nonlinear partial differential equation describes the deterministic movement 
of the population in fitness space over time. 

(Note that, for sufficiently large $N$, the assumptions of Simplifications 1 and 3 
may be violated, so that technically, we can neither expand fitness in $k$ nor 
replace discrete time with continuous time in this regime. Nevertheless, 
\citet{Rouzineetal2003} showed that the continuous equation for the fitness 
distribution, Eq.~\eqref{eq:final-wave-eq}, can be derived in a more general way, 
without assuming $\partial\ln f_k(t)/\partial t \ll 1$, nor expanding fitness in 
$k$, nor neglecting multiple mutations per genome per generation [see the 
transition from Eq.~(1) to Eq.~(11) in the Supplementary Text of the quoted work]. 
In the present work, we use these approximations only to simplify our derivation.)

In the equation above, $\alpha$ and $\sigma$ depend, strictly speaking, on time. 
The dependence is, however, very weak in the limit of a large genome size~$L$. In 
the remainder of this paper, we shall assume that the observation time is very 
small compared to the time in which $\kav$ changes by $L$ and, thus, $\alpha$ and 
$\sigma$ can be considered constant. At the same time, because the distribution is 
very far from the class without deleterious alleles, $|k-\kav|\ll \kav$, a 
considerable shift of the distribution can occur during the observation time.

\subsection{Traveling wave solution}

A broad class of partial differential equations affords solutions in the form of 
traveling waves. A wave is a fixed shape, described by a function $\phi(x)$, that 
moves through space without changing its form. In the case of an evolving 
population, space corresponds to the mutational load, $k$. A traveling wave 
solution to Eq.~\eqref{eq:final-wave-eq} can be expressed as the movement of the 
center of the population over time, $\kav(\tau)$, and the shape of the wave around 
this center, which we write as $\phi[k-\kav(\tau)]$. We therefore make the 
\emph{ansatz} (Simplification~5) that
\begin{equation}\label{eq:ansatz}
  \ln f_k(t) = \phi[k-\kav(\tau)]\,
\end{equation}
After inserting 
Eq.~\eqref{eq:ansatz} into Eq.~\eqref{eq:final-wave-eq} and writing $x=k-
\kav(\tau)$, we obtain
\begin{equation}
\label{eq:wave-shape-equation}
  \sigma x = (1-\alpha)e^{-\phi'(x)} + \alpha e^{\phi'(x)} + v \phi'(x) - 1\,,
\end{equation}
where $\phi'(x)=\partial\phi(x)/\partial x$, and we have introduced the wave 
velocity $v$, that is, the speed of movement of the population center $\kav$ over 
time, $v\equiv\partial\kav(\tau)/\partial\tau$.

For a given wave velocity $v$, Eq.~\eqref{eq:wave-shape-equation} and $\phi(0)$  fully specify the shape of the wave $\phi(x)$. 
Although it is not possible to solve Eq.~\eqref{eq:wave-shape-equation} in 
the general form, analytically, below we will be able to find the velocity of 
the wave.  The first step is to derive $\phi(0)$ from the normalization condition.

\subsection{Normalization condition. Width and speed}

The validity of the approach by \citet{Rouzineetal2003} requires that the {\it logarithm} of the fitness distribution, $\phi(x)$, can be approximated with a smooth function of $x$. In other words, the characteristic scale in $x$ given by the length of the high-fitness tail is much larger than unit.  The condition is always met when population is evolving at many sites at a time. Because  the  fitness distribution itself, $\exp[\phi(x)]$,  is not replaced with a smooth function of $x$, it does not have to be broad: Whether the half-width of the wave is small or large is irrelevant for the validity of the approach. In the treatment of adaptation limit (Section 4), the wave can be either broad or narrow, depending on the range of population sizes. The two cases, however, have different normalization conditions, as follows.

If the wave is narrow,  $\Vark =\text{Var}[x]\ll 1$, most of the 
distribution is localized at $k\approx\kav$, and from the normalization sum $\sum_{k=0}^L{f_k(t)}$ we have 
\begin{equation}\label{eq:phi-of-zero-narrow}
\phi(0)\approx 0. 
\end{equation}
(The exception is the case when $\kav$ is nearly half-integer, and the two adjacent  classes have comparable sizes.)

If the wave is broad, $\Vark \gg 1$,  the normalization condition is less trivial. The normalization sum $\sum_{k=0}^L{f_k(t)}$ can be replaced with an integral, $\int {e^{\phi(x)} dx }$. As we will see momentarily, the main contribution to the integral comes from the interval of $x$ such that $|x|$ is much larger than 1 but much smaller than the high-fitness tail length, so that 
$|\phi'(x)|\ll 1$.  Hence, we can expand the 
exponential terms in Eq.~\eqref{eq:wave-shape-equation} to the first order 
(Simplification~6), and find
\begin{equation}\label{eq:phi-prime}
	\phi'(x) = -\frac{\sigma x}{1-2\alpha-v}\,.
\end{equation}
Clearly, $|\phi'(x)|\ll 1$ when $|x|\ll (1-2\alpha-v)/\sigma$, and therefore this 
latter condition determines the validity of Eq.~\eqref{eq:phi-prime}.
Integrating in $x$, taking into 
account the normalization condition, we obtain the expression for $\phi(x)$ near the wave center
\begin{equation}\label{eq:phi}
  \phi(x) = \ln\sqrt{\frac{\sigma}{2\pi(1-2\alpha-v)}} - \frac{\sigma x^2}{2(1-
2\alpha-v)}\, .
\end{equation}
In particular, the desired expression for $\phi(0)$ has a form
\begin{equation}\label{eq:phi-of-zero}
  \phi(0) = \ln\sqrt{\frac{\sigma}{2\pi(1-2\alpha-v)}}.
  \end{equation}
  
Note that 
$\phi(x)$ was defined as $\phi(x)=\ln f_k(t)$. Therefore, Eq.~\eqref{eq:phi} 
implies that the distribution of genomes in the vicinity of $\kav$ is 
approximately Gaussian, with variance 
\begin{equation}\label{eq:var-k}
  \Vark = (1-2\alpha-v)/\sigma\,.
\end{equation}
Because the variance and the argument of the square root in Eq.~\eqref{eq:phi-of-zero} must be positive, the wave velocity has to satisfy $v<1-
2\alpha$. In other words, if deleterious mutations accumulate, the rate of their accumulation cannot exceed $1-
2\alpha$ if time is measured in units of $U$. Also, as we already stated, the Gaussian expression for the central part of the wave is valid if the width of the wave $\sqrt{\Vark}$ is much larger than 1. For 
moderate or small wave speeds, $|v| \sim 1$ or $|v|\ll 1$, this condition implies 
that $\sigma$ is much smaller than 1. Whether the wave is actually narrow or broad, given the population size and other parameters, will be discussed below for Muller's ratchet and the adaptation regime.

Eq.~\eqref{eq:var-k}, which is known as Fisher Fundamental Theorem, links the width of the population's mutant distribution, 
$\sqrt{\Vark}$, to the velocity $v$ at which the wave is traveling. We emphasize that the validity of this expression is not restricted to the case of broad fitness distribution and can be derived directly from 
Eq.~\eqref{eq:det-expanded-Ualpha} ({\it Appendix}) even if $\Vark\ll 1$. Qualitatively, the theorem states that, the broader the wave, the larger the characteristic difference in fitness between two representative genomes, and the stronger the effect of positive selection on the substitution rate.  

\subsection{High-fitness edge of distribution}

In the previous subsections, we have derived a continuous wave equation that gives 
a deterministic description of how the bulk of the population moves over time in fitness space. However, the speed of the wave remains undefined. As it turns out, the behavior of a small class of best-fit genomes determines the speed.

\begin{figure}
\centerline{\includegraphics[width=3in]{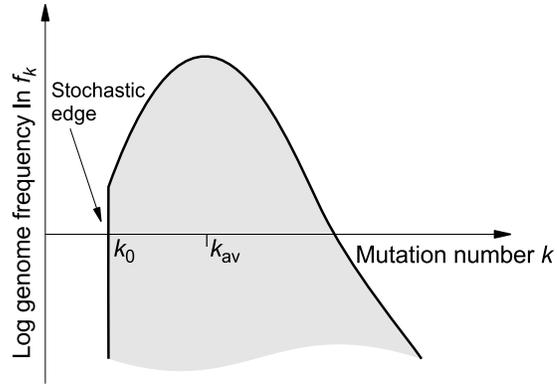}}
\caption{\label{fig:stochastic-edge} Schematic illustration of the solitary wave profile.
The stochastic edge is the minimum-$k$ 
boundary of the population's mutant distribution. There are no genomes with fewer 
than $k_0$ mutations in the population. The speed of the wave is primarily 
determined by how fast mutations are gained or lost at the stochastic edge. 
}
\end{figure}

Below we show that the deterministic wave has a cutoff in the high-fitness tail of the wave whose position is defined by the wave speed and model parameters. The biological reason behind the deterministic cutoff are stochastic factors acting on finite populations. Highly-fit genomes are either gradually lost (Muller's ratchet) or gradually gained (adaptation) by a population.  Thus, the deterministic cutoff is identified with the "stochastic edge" of the wave  that gradually advances or retreats  (Fig.~\ref{fig:stochastic-edge}). The rate of the loss or gain of the least-loaded class, which depends on the population size, ultimately determines the speed of the wave. To obtain the desired condition for the wave speed, in the next subsection, we will match the deterministic description to stochastic behavior of the least-loaded class.

\begin{figure}
\centerline{\includegraphics[width=4in]{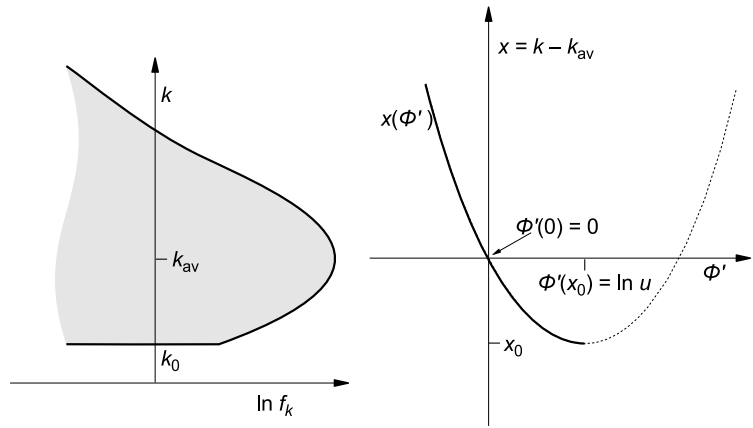}}
\caption{\label{fig:x-of-phi}The minimum of the function $x(\phi')$ determines the 
location of the stochastic edge, $k_0$.
}
\end{figure}

We cannot solve 
Eq.~\eqref{eq:wave-shape-equation} explicitly. Fortunately, 
we can extract the location of the stochastic edge by interpreting 
Eq.~\eqref{eq:wave-shape-equation} as an equation that describes a function 
$x(\phi')$ rather than a function $\phi'(x)$. If the function $x(\phi')$ has an 
absolute minimum, then this minimum implies that $\phi(x)$ has a deterministic cutoff, and thus the minimum must coincide with the stochastic edge 
(Fig.~\ref{fig:x-of-phi}).

Therefore, consider the function $x(\phi') = [(1-\alpha)e^{-\phi'} + \alpha 
e^{\phi'} + v \phi' - 1]/\sigma$. Clearly, $x(\phi')\rightarrow \infty$ for 
$\phi'\rightarrow \pm \infty$, as long as $0<\alpha<1$ or $\alpha=0$ and $v>0$. To 
locate any minima, we calculate the derivative $dx(\phi')/d\phi'$, and find
\begin{equation}\label{dxdphiprime}
  \frac{dx(\phi')}{d\phi'}=[-(1-\alpha)e^{-\phi'} + \alpha e^{\phi'} +v]/\sigma = 
0\,.
\end{equation}
We solve the resulting quadratic equation in $e^{\phi'}$ and arrive at
\begin{equation}\label{eq:u-expression}
  e^{\phi'}=\frac{1}{2\alpha}\Big[-v+\sqrt{v^2+4\alpha(1-\alpha)}\Big]\equiv u\,.
\end{equation}
The alternative solution, with a minus sign in front of the square root, is not 
possible for real-valued $\phi'$. Thus, the derivative of $x(\phi')$ has only a 
single root, which must correspond to an absolute minimum. The value of $x(\phi')$ 
at this minimum, which we denote by $x_0$, follows as
\begin{equation}\label{eq:x_zero}
	x_0 = -\frac{1}{\sigma}(1-2\alpha u-v\ln u - v)\,,
\end{equation}
where we have made use of the identity $(1-\alpha)/u=\alpha u+v$. For any mutation 
classes $k$ corresponding to $x<x_0$, we have $f_k\equiv0$. For the continuous 
approach to work, we need $|x_0|\gg 1$ (Simplification~4).

\subsection{Difference between fitness distribution at center and edge}

Our eventual goal is to find an expression that relates the speed of the wave, $v$, to the 
population size $N$, mutation rate $U$, selective coefficient $s$, and the 
fraction of beneficial mutations $\alpha$. It turns out that we can achieve this 
goal by considering the difference $\phi(0)-\phi(x_0)$. By evaluating this 
difference in two alternative ways, we end up with the desired relation.

First, we note that
\begin{equation}\label{eq:delta-phi}
  \phi(0)-\phi(x_0) = \int_{x_0}^0 \phi'(x) dx\,.
\end{equation}
We can evaluate the integral on the right-hand side by making the subsitution 
$x=x(\phi')$, integrating by parts, and using $\phi'(x_0)=\ln u$ and $\phi'(0)=0$. 
(The former condition stems from the definition of $u$, while the latter condition 
is a consequence of Eq.~\eqref{eq:phi-prime}; see also Fig.~\ref{fig:x-of-phi}.) 
We arrive at
\begin{equation}\label{eq:delta-phi2}
  \phi(0)-\phi(x_0) = -x_0\ln u - \int_{\ln u}^0 x(\phi') d\phi'\,.
\end{equation}
We can evaluate the remaining integral on the right-hand side by integrating over 
Eq.~\eqref{eq:wave-shape-equation}\,. We obtain, after inserting 
Eq.~\eqref{eq:x_zero} into the first term on the right-hand side of 
Eq.~\eqref{eq:delta-phi2},
\begin{equation}\label{eq:delta-phi3}
  \phi(0)-\phi(x_0) = \frac{1}{\sigma}\Big(1-2\alpha -\frac{v}{2}[\ln^2(eu)+1]-
2\alpha u \ln u\Big)\,,
\end{equation}
where $e$ is Euler's constant, $\ln(e)=1$.

We have now calculated $\phi(0)-\phi(x_0)$ by integrating over $\phi'(x)$. 
Alternatively, we can calculate this difference by evaluating $\phi(x)$ directly 
at 0 and at $x_0$. $\phi(0)$ follows from Eq.~\eqref{eq:phi-of-zero}
if the argument of the logarithm is much smaller than 1 (large half-width of the 
wave), and $\phi(0)=0$ in the opposite limit (small half-width). The quantity 
$\phi(x_0)$ represents the logarithm of the genome frequency right at the 
stochastic edge. Since this value is dominated by stochastic effects, we cannot 
evaluate $\phi(x_0)$ on the basis of the deterministic theory we have developed so 
far. Therefore, we proceed as follows. We give the genome frequency at the 
stochastic edge proper stochastic treatment and base our estimate of the expected 
genome frequency at the stochastic edge on these probabilistic arguments.

\subsection{Stochastic treatment of the least-loaded class}
\label{stoch-treatment}
The deterministic derivation in the previous subsections demonstrates the existence of a continuous 
set of solitary waves with various speeds, i.e., average substitution rates. Now, to choose the 
correct solution and determine the actual substitution rate, we have to take into 
account stochastic effects acting on the least-loaded class, $k=k_0$. At this point, finite 
population size enters the scene. We describe the dynamics of the least-loaded 
class using the one-site, two-allele model (see reviews in \citet{Kimura64}, 
\citet{Rouzineetal2001}, or \citet{Ewens2004}). The frequency of the least-loaded class is analogous to the concentration of the better-fit allele. To justify the deterministic 
treatment of the next-loaded class (and other classes) and estimate the error introduced by that approximation, we will estimate its size in the limits of adaptation and Muller's ratchet in Sections 3 and 4, respectively.

The accurate treatment of the stochastic dynamics of least-loaded class, $f_{k_0}(t)$, requires the use of a diffusion equation that includes both deterministic factors (selection and mutation) and random genetic drift. For the purpose of the present work, a good accuracy can be ensured by a simpler approach, as follows.

A well-known fact following from the diffusion equation is that deterministic factors dominate when the frequency of a minority allele in a population (in our case, $f_{k_0}(t)$) is much larger than $1/(N|S|)$ \citep{MaynardSmith71,Barton95,Rouzineetal2001}; in the opposite case, random drift rules, and the minority is most likely to be lost. Here $\exp(|S|)$ is the fitness advantage of the better-fit allele. If the best-fit class is much larger than that value, we can use the deterministic equation,  Eq.~\eqref{eq:det-expanded-Ualpha}, which yields
\begin{equation}\label{eq:bar-fkz-diffeq}
\partial f_{k_0}/\partial t = M(t)-S f_{k_0}(t)
\end{equation}
with $M(t)=U\alpha f_{k_0+1}(t)$ and $S=U+s(k_0-k_\text{av})$. The parameter $S$ 
represents the effective coefficient of selection against the best-fit class in 
the population. It consists of two parts: the positive part $U$ due to mutations 
removing genomes from the class, and the negative part due to the fitness difference between 
the least-loaded sequence and the average genome, $s(k_0-k_\text{av})$. 

As we show 
below, $S$ is positive and $M$ is negligible in the ratchet limit, $v>0$, and negative in the adaptation 
limit, $v<0$, where $M(t)$ is more pronounced but still relatively small. In the ratchet limit, the least-loaded class decreases exponentially until it hits the characteristic size $1/(N|S|)$ and is lost. In the adaptation limit, the least loaded class is created due to a mutation and becomes established, i.e., survives and grows further with probability on the order of $1/2$, when its frequency in a population exceeds $1/(N|S|)$. The cost of treating a characteristic frequency $1/(N|S|)$ as a sharp threshold is an undefined numeric factor multiplying the population size, $N$. Because the speed of evolution, as it comes out in the end, depends on $N$ logarithmically, and the interesting values of $N$ are quite large, the error is relatively small. 

\begin{figure}
\centerline{\includegraphics[width=4in]{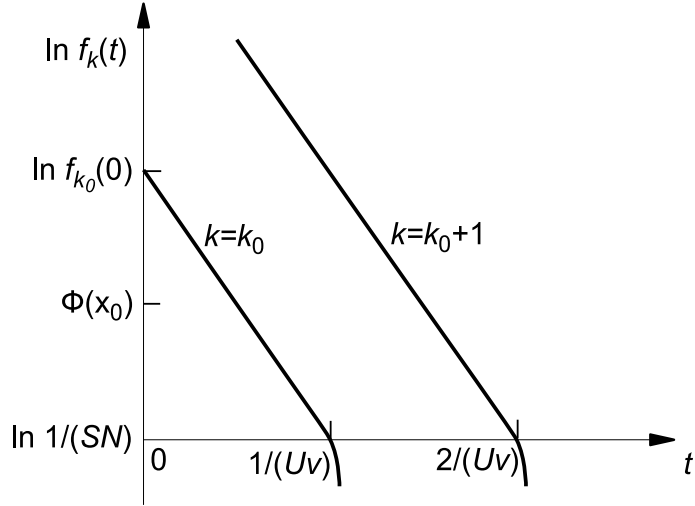}}
\caption{\label{fig:loss-of-best-class} 
Schematic illustration of the loss of best-fit class.
The frequency of the least-loaded class, 
$\fkz(t)$, decays approximately exponentially in time, until it reaches the 
stochastic threshold $1/(SN)$. At the stochastic threshold, the best-fit class is 
rapidly lost (in time $\sim 1/S$).}
\end{figure}

The deterministic formalism described in the previous section predicts a monotonous dependence on time for a fitness class (the time derivative changes sign only at $k=k_{\rm av}$). In fact, because the mutational load of the least-loaded class $k_0$ changes abruptly in time, the size of the current least-loaded class oscillates in time producing a saw-like dependence. In the case of the Muller's-ratchet, the class contracts until lost (Fig. \ref{fig:loss-of-best-class}), and the next-loaded class takes it place, and in the case of adaptation, the class expands until beneficial mutations create a new least-loaded class (Fig. \ref{fig:gain-of-new-class}). To match the two formalisms, we require that the logarithm of the solitary wave at the edge, $\phi(x_0)$, is equal to the value $\ln f_{k_0}(t)$ averaged over one period of oscillations, as given by
\begin{equation}
\label{match}
\phi(x_0)=\frac{1}{2}\left[ \ln{\frac{1}{N|S|}}+\ln f_\text {max}\right]
\end{equation}
where $f_\text {max}$ is the maximum size of the least-loaded class size. The value of $f_\text {max}$ is given by $f_{k_0}(t)$ at the moment $t$ when either the previous least-loaded class was just lost (ratchet case) or a new least-loaded class was just established (adaptation case). In the next two sections, we derive the value of $f_\text {max}$ and determine $\phi(x_0)$ from Eq.~\eqref{match} in each case separately. To partly compensate the error introduced by discreteness of $k$ and matching the continuous fitness distribution to the oscillating edge, below we derive corrections to the continuous approximation. In the case of adaptation, the uncompensated error in the substitution rate amounts to 10-20\% in the entire parameter range we tested (see simulation in Section 4).

We note that the time dependence of the classes other than the least-loaded class also has an oscillating component. The period of these oscillations is the same as for the least loaded class, corresponding to a shift of the wave by one notch in $k$. Because the average class size increases exponentially away from the edge, the relative magnitude of oscillations decreases rapidly away from the edge, and the described method of matching the edge to the bulk yields fair accuracy (see simulation in the next two sections). The remaining effect of oscillations will be partly accounted for by the correction for discreteness of $k$, which we introduce in the next two sections.

\section{Muller's ratchet}

In the absence of beneficial mutations, all genomes have at least as many mutations as 
their parents. Therefore, if the class of mutation-free genomes is lost from the 
population because of genetic drift, it cannot be regenerated in an 
asexual population. Moreover, now the class of one-mutants is at risk of being 
lost, and once it is lost, the class of two-mutants is at risk, and so on. In this 
way, an asexual population will inexorably experience accumulation of mutations 
and decay of fitness. This process was first described by \citet{Muller64} in an 
argument for why sexual reproduction is beneficial, and is commonly referred to as 
Muller's ratchet \citep{Felsenstein74}.

In our study, the Muller's-ratchet regime corresponds to the limit of high average fitness, $\alpha\rightarrow 0$. In this case, beneficial mutations play a negligible role, and the speed $v$ 
of the wave is positive, $0<v<1$.  The following derivation is valid if $s\ll U$, which condition ensures  that the high-fitness tail of the distribution is long. Incidentally, the same condition implies a large half-width of the wave. (Note that the two conditions differ in the case of adaptation, see the next section). We consider the case of moderately large population sizes, $\ln N < U/s$ (a more accurate condition is given in Appendix A.1), when clicks of the ratchet are sufficiently frequent, so that the fitness distribution never equilibrates and moves almost continuously. In the opposite case of very large population sizes,,  ratchet clicks are rare, 
and the population can equilibrate between subsequent ratchet clicks. The ratchet rate is exponentially small and is estimated from the average time the fittest class takes 
to drift from its equilibrium value to an occupancy of zero \citep{Haigh78,Stephanetal93,GordoCharlesworth2000a,GordoCharlesworth2000b}.

\subsection{Solitary wave approach in the ratchet limit}

Before we discuss the treatment of the 
stochastic edge in this case, we simplify the deterministic part of our theory, 
that is, the right hand side of Eq.~\eqref{eq:delta-phi3}. 

From the expression for $u$, Eq.~\eqref{eq:u-expression}, we find for $\alpha=0$:
\begin{equation}\label{eq:u-expand-MR}
  u = 1/v\,.
\end{equation}
With this expression, we obtain from Eq.~\eqref{eq:x_zero}
\begin{equation}\label{eq:x-zero-MR}
  x_0 =- \frac{1}{\sigma}[1-v\ln(e/v)]\,.
\end{equation}
Similarly, Eq.~\eqref{eq:delta-phi3} becomes
\begin{equation}\label{eq:delta-phi-MR}
  \phi(0)-\phi(x_0)=\frac{1}{\sigma}\Big(1-\frac{v}{2}\Big[\ln^2 
\frac{e}{v}+1\Big]\Big)\,.
\end{equation}

As already stated, the approach is valid if the high-fitness tail is long, $|x_0|\gg 1$ (Simplification~4), which reduces to the condition $\sigma \ll 1$. Thus, our treatment does not predict Muller's ratchet in a broad parametric interval unless the selection coefficient is much smaller than the 
total mutation rate. In this case, the half-width of the wave is also large, and $\phi(0)$ in the left-hand side of Eq.~\eqref{eq:delta-phi-MR} is given by 
\begin{equation}\label{eq:phi-zero-MR}
\phi(0) = \ln\sqrt{\frac{\sigma}{2\pi(1-v)}}\,,
\end{equation}
which represents Eq.~\eqref{eq:phi-of-zero} with $\alpha=0$. 

We will now derive an expression for $\phi(x_0)$ based on the stochastic treatment 
of the least-loaded class. Because $\alpha=0$ in the Muller's ratchet limit, we 
set $M(t)\equiv 0$ in Eq.~\eqref{eq:bar-fkz-diffeq}. Further, from $S=U+s(k_0-
\kav)=U(1+\sigma x_0)$, we obtain
\begin{equation}\label{eq:S-MR}
  S=Uv\ln(e/v)\,,
\end{equation}
where we have made use of Eq.~\eqref{eq:x-zero-MR}; we have $S>0$ at all $v<1$. 
After integrating Eq.~\eqref{eq:bar-fkz-diffeq}, we 
find that the expected frequency of the least-loaded class decays as
\begin{equation}\label{eq:fkz-of-t-MR}
  \fkz(t) = \fkz(0) e^{-St}\,,
\end{equation}
where $t=0$ is the point in time right after the previously least-loaded class 
$k_0-1$ has been lost.  Eq.~\eqref{eq:fkz-of-t-MR}  applies only while  $ \fkz(t)$ is much higher than 
the stochastic threshold $1/(NS)$. When the stochastic threshold is reached, random genetic drift 
overcomes effects of selection,  and the 
$k_0$ class is lost. Note that the stochastic threshold is defined within the accuracy of a numerical coefficient on the order of 1.

The next step is to couple the dynamics of the least-loaded class to 
the movement of the wave. According to the approach outlined in Section~\ref{stoch-treatment}, we match the edge value  $\phi(x_0)$ to the average $\ln\fkz(t)$ during the time in which the 
least-loaded class remains above the threshold condition. Using Eq.~\eqref{match} with $f_\text{max}=\fkz(0)$, we obtain
\begin{equation}\label{eq:phi-of-x-ave-MR}
 \phi(x_0) = \frac{1}{2}\Big[\ln \fkz(0) + \ln\frac{1}{SN}\Big] \,.
\end{equation}

As per the definition of $v$, the wave moves one notch in time 
$t_\text{click}=1/(Uv)$. The time until the least-loaded class reaches the 
stochastic threshold condition, $t_\text{loss}$, follows from 
Eq.~\eqref{eq:fkz-of-t-MR} as
\begin{equation}
  \frac{1}{SN} = \fkz(0) e^{-St_\text{loss}}\,.
\end{equation}
After equating $t_\text{click}$ and $t_\text{loss}$, we find
\begin{equation}\label{eq:fkz-MR2}
  \ln \fkz(0) = \frac{S}{Uv} + \ln \frac{1}{SN}\,.
\end{equation}
We now insert Eq.~\eqref{eq:fkz-MR2} into Eq.~\eqref{eq:phi-of-x-ave-MR} and 
obtain with Eq.~\eqref{eq:S-MR}
\begin{equation}\label{eq:phi-of-x-MR2}
  \phi(x_0)=-\ln[NUv^{3/2}\ln(e/v)]\,.
\end{equation}
Because the stochastic threshold $1/(NS)$ is an estimate defined 
up to a numerical coefficient not to exceed the accuracy of our calculation, we 
set the numerical coefficient inside of the logarithm  to 1. The missing numerical 
coefficient introduces a small error, because we have assumed that the population 
size and, hence, the argument of the logarithm are large. 

Previously, \citet{Rouzineetal2003} matched $\phi(x_0)$ directly to the stochastic threshold $\ln[1/(NS)]$, which leads to a slight underestimation of  $\phi(x_0)$. The only difference from the old treatment is an additional factor of $v^{1/2}$ in the argument of the outer logarithm in Eq.~\eqref{eq:phi-of-x-MR2}.

We neglected the stochastic effects acting on the next-loaded class, $k=k_0-1$, on 
the grounds that its size is larger than then the size of the least-loaded class, which is on the order or larger than the stochastic threshold; hence, the random drift term in the two-allele diffusion equation for the next-loaded class can be neglected as compared to the selection term. 
The average ratio of the two sizes can be estimated, in the deterministic fashion, 
from the logarithmic slope of the fitness distribution, as given by $f_{k_0-1}/ 
f_{k_0}=\exp[\phi'(x_0)]=u=1/v$, Eq.~\eqref{eq:u-expand-MR}. Thus, in the typical 
case when $v$ is not close to 1, the next-loaded class is, at least, several-fold 
larger than the least-loaded class. The error introduced by the 
approximation is, again, only a numeric factor at $N$, Eq. \eqref{eq:phi-of-x-MR2}. Computer simulation 
confirms the small numeric effect of the error (see below).

\subsection{Correction for discontinuity at $k=k_0$}

In the derivation of the wave equation, we made the assumption that $\ln f_{k+1}-
\ln f_k$ can be approximated by $\partial \ln f_k/\partial k$ (Simplification~4). 
This approximation is valid if the first derivative of $\ln f_k$ does not change 
much from $k$ to $k+1$, that is, if $\ln f_k$ is approximately linear between $k$ 
and $k+1$. We can state this condition more formally by demanding that
\begin{equation}\label{continuity0}
  \Big|\frac{\partial \ln f_k}{\partial k}\Big| \gg \Big|\frac{\partial^2 \ln 
f_k}{\partial k^2}\Big|\,.
\end{equation}
With $\partial \ln f_k/\partial k= \phi'$, we can rewrite this condition as
\begin{equation}\label{continuity}
  |\phi'|\Big|\frac{dx}{d\phi'}\Big|\gg 1\,.
\end{equation}
For $v \sim 1$ and $|x|\sim|x_0|$, from Eq.~\eqref{dxdphiprime} at $\alpha=0$ and 
Eq.~\eqref{eq:u-expand-MR}, we estimate $\phi'(x)\sim\phi'(x_0)=\ln(u)\sim 1,\ 
|dx/d\phi'|\sim1/\sigma$. Thus, for a fitness class somewhere in the middle of the 
tail and $v\sim 1$, the continuity condition Eq.~\eqref{continuity} is equivalent 
to $\sigma\ll 1$. Therefore, the tail must be long, $|x_0|\gg 1$, 
Eq.~\eqref{eq:x-zero-MR}, as we already mentioned several times on intuitive grounds. However, 
near the fitness edge $x=x_0$, we have $dx/d\phi'=0$, because we defined $x_0$ to 
be the minimum of $x(\phi')$. Therefore, the condition is violated very close to 
the edge. This effect is especially important at $v\approx 1$, where the 
condition on $\sigma$ becomes more restrictive. Indeed, the tail length vanishes 
at $v\rightarrow 1$, Eq.~\eqref{eq:x-zero-MR}.

Condition \eqref{continuity} is trivially violated in a narrow vicinity of the wave center, $x=0$, where $\phi'(0)=0$. Because the first derivative in the left-hand side of Eq. ~\eqref{continuity0} is zero, a more appropriate condition of continuity in that region is that the second derivative is much larger than the third derivative, which is met at $\sigma \ll 1$.

\begin{figure}
\centerline{\includegraphics[width=4in]{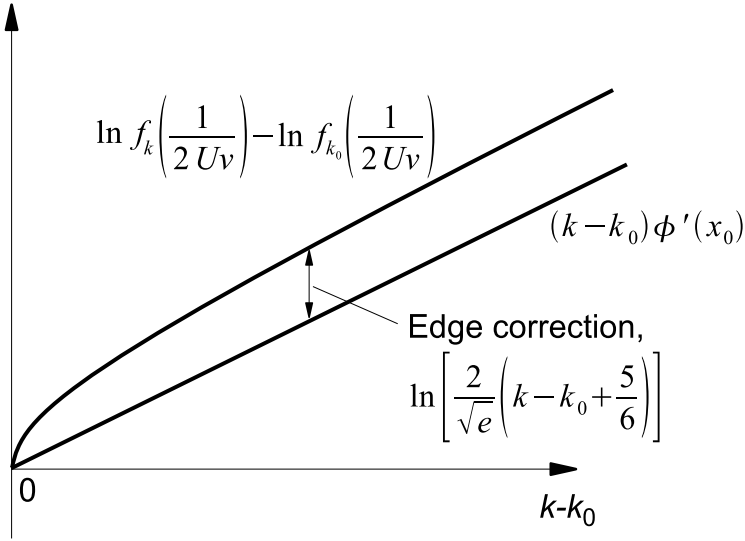}}
\caption{\label{fig:edge-correction} 
Schematic illustration of correction for discreteness.
Our continuous treatment of fitness classes 
predicts that $\ln f_k(t)-\ln \fkz(t)$ grows linearly in $k-k_0$ near the high-
fitness edge, $k-k_0\ll |x_0|$. Discreteness of $k$ introduces a correction near 
the edge, approximately equal to $\ln[(2/\sqrt{e})(k-k_0+5/6)]$.}
\end{figure}

Right at the edge, $\ln f_k$ grows faster with $k$ than our continuity 
approximation allows for. The inflation of the $\ln f_k$ values for $k$ close to 
$k_0$ spreads inwards towards higher $k$ values in a deterministic fashion (see 
Fig.~\ref{fig:edge-correction}). We can investigate the effect of this 
perturbation by considering the discrete balance equations near the edge,
\begin{align}\label{eq:balance-near-edgeI}
  \frac{d\fkz}{dt}&= -S\fkz\,,\\
  \frac{df_k}{dt} &= -Sf_k + Uf_{k-1}\,,\qquad k\geq k_0+1\,,
  \label{eq:balance-near-edgeII}
\end{align}
with $S=Uv\ln(e/v)$. These equations follow directly from 
Eq.~\eqref{eq:det-expanded-Ualpha} if we approximate $S$ with its value at the edge, as given by 
Eq.~\eqref{eq:S-MR}. As we show in the appendix, this set of differential 
equations has a solution under periodic initial conditions of the form $f_{k-
1}(0)=f_k[1/(Uv)]$. In other words, under the stationary process, the wave moves 
one notch in time $1/(Uv)$, periodically repeating its shape near the edge.

Note that Eqs.~\eqref{eq:balance-near-edgeI} and~\eqref{eq:balance-near-edgeII} 
apply only close to the edge, because we have assumed that $S$ is constant in $k$. 
Far from the edge, we have to use instead a more general equation, which we solve 
in the continuous approximation as described above.

Our continuous approximation predicts that the difference between $\ln f_k$ and 
$\ln f_{k_0}$ near the high-fitness edge, $k-k_0\ll |x_0|$, is given by the linear 
expression $\partial \ln f_k/\partial k \approx\phi'(x_0)(k-k_0)=\ln(1/v)(k-k_0)$, 
where the last equality follows from Eq.~\eqref{eq:u-expand-MR}. However, when we 
integrate the set of differential equations Eqs.~\eqref{eq:balance-near-edgeI} 
and~\eqref{eq:balance-near-edgeII}, and compare the values of $\ln f_k(t)$ and 
$\ln \fkz(t)$, for example, in the middle of one cycle, $t=1/(2Uv)$, we find 
[Eq.~\eqref{eq:comparison1} in the appendix]
\begin{equation}
\ln f_{k}\Big({1\over2Uv}\Big)-\ln f_{k_0}\Big({1\over2Uv}\Big)\approx
\ln(1/v)(k-k_0)+\ln\Big[{2\over\sqrt{e}}\Big(k-k_0+{5\over6}\Big)\Big]\,.
\end{equation}

In other words, on the right hand side of Eq.~\eqref{eq:delta-phi-MR}, which 
corresponds to our estimate of the difference $\phi(0)-\phi(x_0)$ from the 
continuous approximation, we have to add a correction term of the form 
$\ln[(2/\sqrt{e})(|x_0|+5/6)]$ to account for the fact that, given $\phi(x_0)$, 
the value of $\phi(0)$ is slightly larger than what our continuous approximation 
predicts. Note that this correction term does not depend on any model parameters. With $x_0$ given by Eq.~\eqref{eq:x-zero-MR}, the 
correction term becomes
\begin{equation}\label{eq:cont-correction}
  \ln\Big[\frac{2}{\sqrt{e}}\Big(1-v\ln\frac{e}{v}+\frac{5\sigma}{6}\Big)\Big]-\ln 
\sigma\,.
\end{equation}

Combining the correction term with Eqs.~\eqref{eq:delta-phi-MR}, 
\eqref{eq:phi-zero-MR},~and~\eqref{eq:phi-of-x-MR2}, and dropping all the numerical constants 
multiplying $N$ inside the logarithm, we arrive at our final result 
\begin{equation}\label{finalratchet}
\sigma\ln(NU\sigma^{3/2})\approx \Big[1-
\frac{v}{2}\Big(\ln^2\frac{e}{v}+1\Big)\Big]
         - \sigma\ln\left[\sqrt{\frac{v^3}{1-v}} \frac{\ln(e/v)}{1-
v\ln(e/v)+5\sigma/6}\right]\,.
\end{equation}
This expression relates the selection strength $\sigma =s/U$, population size $N$, 
and mutation rate $U$ to the normalized ratchet rate, $v=(1/U)d\kav/dt$. We can 
evaluate this expression in two ways. First, we can assume the point of view that 
Eq.~\eqref{finalratchet} describes $N$ as a function of $v$ and model parameters. 
Thus, we can directly plot $N$ over the entire range of $v$ ($0<v<1$). The solid 
lines in Fig.~\ref{fig:ratchet} were generated in this way. Alternatively, we can 
solve Eq.~\eqref{finalratchet} numerically for $v$ at a given value of $N$.

We note that, at very small $\sigma$ and large $N$, the second term in the right-
hand side of Eq.~\eqref{finalratchet}, originating from the stochastic edge 
treatment and the correction for discontinuity, can be neglected. In this limit, 
the substitution rate normalized to the mutation rate is expressed in terms of a 
single composite parameter $\sigma\ln(NU\sigma^{3/2})$. When this parameter is 
equal to 1, the ratchet rate, in this limit, becomes exactly zero. In fact, the 
ratchet speed is predicted \citep{GordoCharlesworth2000a} to be finite albeit 
exponentially small at $\sigma\ln(NU\sigma^{3/2})>1$, when loss of the fittest 
genotype occurs so infrequently that the solitary wave approach breaks down (see 
below). At realistically small $\sigma$, the second term in 
Eq.~\eqref{finalratchet} is important, as we show below by simulation, especially 
near $v=1$, where the first term vanishes as $(1-v)^3$, and at very small $v$, 
where the second term smears out the critical point $\sigma\ln(NU\sigma^{3/2})=1$.

The selection coefficient $s$ enters Eq.~\eqref{finalratchet} through its rescaled 
version $\sigma=s/U$. Likewise, the ratchet speed $-V=dk_{\rm av}/dt$ enters in 
the rescaled form, $v=-V/U$. On first glance, this scaling looks unusual in 
comparison to the standard diffusion limit. In this limit, all results are 
expressed in terms of time rescaled with the population size, $t'=t/N$ 
\citep{Ewens2004}, and rescaled mutation rate and selection coefficient, $U'=NU$ 
and $s'=Ns$. However, it is straightforward to express Eq.~\eqref{finalratchet} in 
terms of the standard diffusion limit scaling. Note that a rescaling of time with 
$N$ means that we also have to rescale the ratchet rate, $V'=NV$, and note further 
that $v=-V/U=-V'/U'$ and $\sigma=s/U=s'/U'$. Hence, if we replace the factor $NU$ 
on the left-hand side of Eq.~\eqref{finalratchet} with $U'$, 
Eq.~\eqref{finalratchet} is given entirely in terms of the rescaled quantities of 
the standard diffusion limit, $V',\ s'$ and $U'$.

\subsection{Comparison with simulation results and other studies}

\begin{figure}[t]
\centerline{\includegraphics[width=6in]{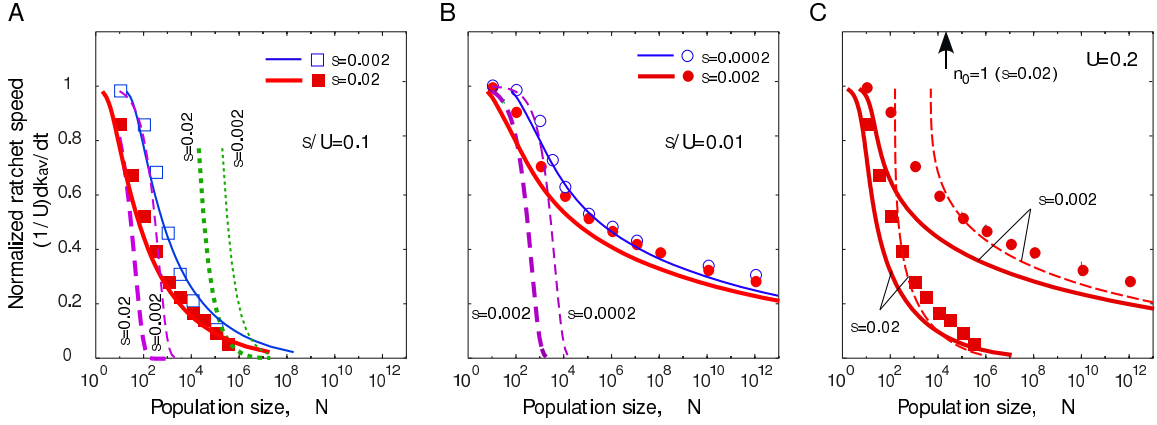}}
\caption{\label{fig:ratchet}
Normalized ratchet speed as a function of population size: analytic results versus simulation.
Symbols correspond to 
simulation results, and lines correspond to theoretical predictions. The 
simulation results were obtained as described \citep{Rouzineetal2003}.  Beneficial 
mutations are absent ($\alpha=0$). (A) Results for $\sigma=s/U=0.1$. The solid 
blue and red lines follow from the present work, Eq.~\eqref{finalratchet}. The 
dotted green lines follow from \citet{GordoCharlesworth2000a}, Eqs.~(3a) and (3b). 
The dashed purple lines follow from \citet{Lande98}, Eq.~(2c) times $NU$. 
Parameters are shown. (B) As (A), but for $\sigma=s/U=0.01$. The dashed purple 
lines follow again from \citet{Lande98}; the equivalent of the green lines in (A) 
falls outside of the axis range. (C) Effect of discreteness correction. The red 
squares and circles are identical to those in (A) and (B). The solid lines 
correspond to the prediction of traveling-wave theory without the discreteness 
correction, i.e., without a factor of $\sigma$ inside of the logarithm on the 
left-hand-side of Eq.~\eqref{finalratchet}, and without the denominator $1-
v\ln(e/v)+5\sigma/6$ inside the logarithm on the right-hand side of 
Eq.~\eqref{finalratchet}. The dashed lines correspond to Eq.~\eqref{finalratchet} 
without the entire second term on the right-hand-side. The arrow shows the value 
of $N$ at which the size of the least-loaded class $n_0=Ne^{-U/s}$ \citep{Haigh78} 
in the equilibrium distribution is 1 (at $s/U=0.1$).
}
\end{figure}

We carried out simulations of Muller's ratchet as described 
\citep{Rouzineetal2003}, and compared the measured ratchet speed to the speed 
predicted by Eq.~\eqref{finalratchet} (Fig.~\ref{fig:ratchet}). As shown 
previously \citep{Rouzineetal2003}, the traveling-wave theory leads to accurate 
predictions of the ratchet rate over a wide range of population sizes. Even though 
the theory is derived under the assumption that $N$ is large, we find that our 
prediction for the ratchet rate is accurate already for population sizes as small 
as $N=10$, and continues to be accurate throughout the entire range of 
biologically reasonable population sizes, until $\sigma\ln(NU\sigma^{3/2})$ 
becomes larger than 1. Thus, our results connect expressions for the ratchet rate 
derived by \citet{Lande98} or \citet{GordoCharlesworth2000a} that work, 
respectively, for either very small or very large population sizes.

The transition to the regime of large $N$ occurs when the size of the least-loaded 
class is large, $N\exp(-1/\sigma)\gg 1$ \citep{GordoCharlesworth2000a} (Fig. 5c). 
Then, clicks of the ratchet are rare, and the population is at equilibrium most of 
time. In contrast, in the intermediate interval of $N$ we study, the least-loaded 
class is not sufficiently large and is frequently lost. Therefore, a population 
does not have time to equilibrate between clicks. The non-equilibrium nature of 
the ratchet is witnessed by an almost constant, non-zero effective selection 
coefficient of the least-loaded class, $S > 0$ and the fact that the width of the 
distribution is smaller than the equilibrium width, $\sqrt{\Vark}<\sqrt{U/s}$, 
Eq.~\eqref{eq:var-k}. The transition to the regime of small $N$ \citep{Lande98} 
occurs at $Ns\sim 1$, when fixation events of deleterious mutations at different 
sites are separated in time, and one-site theory applies.

We found that the term correcting for the discontinuity at $k=k_0$, 
Eq.~\eqref{eq:cont-correction}, substantially improved the agreement between the 
numerical simulations and the analytical prediction of the ratchet speed for a 
large interval of population sizes  (Fig.~\ref{fig:ratchet}C). While the 
traveling-wave approximation (based on continuous treatment of fitness classes) 
combined with the stochastic treatment of the highest-fitness class is good enough 
to make useful predictions about the speed of Muller's ratchet, the discreteness 
correction is necessary to achieve high prediction accuracy. The correction term 
becomes irrelevant only for very large population sizes, where the ratchet speed 
is dominated by the first term on the right-hand side of Eq.~\eqref{finalratchet} 
(Fig.~\ref{fig:ratchet}C, dashed lines).

Note that \citet{Rouzineetal2003} also used a correction term to account for the 
discontinuity at $k=k_0$. However, their term was obtained by fitting an 
interpolation formula to the numerical solution of Eqs.~
\eqref{eq:balance-near-edgeI} and~\eqref{eq:balance-near-edgeII}. By contrast, here we derived the 
correction term analytically, as the asymptotic behavior near the high-fitness 
edge of the wave.

\section{Speed of adaptation}

If a population experiences a sudden change in environmental conditions or is 
introduced into a new environment, it will initially be ill-adapted. Over time, 
the population will accumulate beneficial mutations until a new mutation-selection 
balance is reached. The process of accumulating beneficial mutations is called 
\emph{positive adaptation}, and the rate at which fitness changes over time is the 
\emph{speed of adaptation}. 

In our model, in the case of positive adaptation the speed of the wave is 
negative, $v<0$, because the wave moves towards smaller mutation numbers $k$. 
Furthermore, we make the assumption that we are far from mutation--selection balance, $|v|\gg1$, i.e., $|d\kav/dt|\gg U$. In this regime, the rate of deleterious mutations is a small correction 
that can be neglected, and only the beneficial mutation rate affects our results. 
Therefore, it is more convenient to introduce, instead of $v$, the average substitution rate of 
beneficial mutations $V$, as given by
\begin{equation}
  V=-\frac{d\kav(t)}{dt} = -Uv,
\end{equation}
and the beneficial mutation rate $\Ub = \alpha U$. The derivation that follows applies if $s$ is much larger than $U_b$, and the population size is sufficiently large, so that $V\gg s/\ln(s/U_b)$. The latter condition ensures that the high-fitness tail of the distribution is long and the whole approach is valid. Within this validity region, we will consider two cases: moderate substitution rates, $ s/\ln(s/U_b)\ll V\ll s$,
where the half-width of the fitness distribution is small, and relatively high substitution rates, $ V\gg s$, where the distribution is broad.

\subsection{Solitary wave approach in the adaptation limit}

We expand $u$, $x_0$, 
$\phi'(x_0)$ [Eqs.~\eqref{eq:u-expression} and~\eqref{eq:x_zero}] and the right-hand side of Eq.~\eqref{eq:delta-phi3} for large negative $v$ (dropping all terms 
small compared to $|v|$), and find
\begin{align}\label{eq:u-lnu-AE}
  u &= \frac{V}{\Ub}\,,\quad \phi'(x_0) =\ln u= \ln \frac{V}{\Ub}\,,\\
  x_0 &= -\frac{V}{s}\Big(\ln\frac{V}{\Ub}-1\Big)\,,\label{eq:x-zero-AE}
\end{align}
and
\begin{equation}\label{eq:phi-zero-phi-x-zero-AE}
  \phi(0)-\phi(x_0) = \frac{V}{2s}\Big(\ln^2\frac{V}{e\Ub}+1\Big)\,.
\end{equation}

The approach rests on smoothness of $\ln f_k$. We obtain the continuity condition for $\ln f_k$ as before from Eq.~\eqref{continuity}, using $\phi'\sim \ln u = \ln(V/\Ub)$ and 
Eq.~\eqref{dxdphiprime}. In Eq.~\eqref{dxdphiprime}, we keep only the third term 
in brackets: the first term is small, and the second term is important only near 
the edge. The resulting condition, $V\gg s/\ln(V/\Ub)$, is equivalent to $|x_0|\gg 
1$. Just as in the case of Muller's ratchet, the high-fitness tail must be long 
for our approach to work (Simplification~4). 

Unlike in the ratchet case, even though the high-fitness tail is long, the half-width of the wave is not necessarily large. Large adaptation rates (large population sizes) correspond to a broad distribution, $\text{Var}[k]\gg 1$, and intermediate adaptation rates (intermediate population sizes) correspond to a narrow distribution, $\text{Var}[k]\ll 1$. For the two respective cases, from Eqs.~\eqref{eq:phi-of-zero} and \eqref{eq:phi-of-zero-narrow} with $V\gg U$ ($v<0, |v|\gg 1$), we obtain
\begin{eqnarray}\label{eq:phi-zero-AE}
  \phi(0) &=& -\frac{1}{2} \ln\frac{2\pi V}{s},\ \ V\gg s \\
  \label{eq:phi-zero-AE-narrow}
  \phi(0) &=& 0,\ \ s/\ln(V/\Ub)\ll V\ll s.
\end{eqnarray}

Again, we have to couple the deterministic wave equation to the stochastic 
behavior of the least-loaded class following the approach in Section 2.7. The condition that only the best-fit class should be treated stochastically reads $ V\gg U_b$, which is equivalent to $s|x_0| \gg U_b$ (Simplification 7 and the end of this subsection). 

Far above the stochastic threshold, the dynamics of the least-loaded class follows Eq.\eqref{eq:bar-fkz-diffeq} with the effective
selection coefficient $S$ and the mutation term $M$ given by
\begin{equation}\label{SandM}
S=U+sx_0 \approx -V\ln(V/\Ub),\ \ M=\Ub f_{k_0+1}.
\end{equation}
Thus, the effective selection coefficient is negative, causing exponential expansion of the class.
Unlike in the ratchet case, beneficial mutation events cannot be neglected: Their role is to add more and more clones to the class, resulting in a
time-dependent prefactor influencing the growth of the class on large time scales. However, the logarithmic derivative of that growth is primarily determined by selection alone. 
To demonstrate the validity of this statement, we evaluate the relative magnitude 
of the two terms $M$ and $S\fkz$ in Eq.\eqref{eq:bar-fkz-diffeq}. From Eq.~\eqref{SandM}, we have 
\begin{equation}
 \frac{|S|\fkz}{M}=\frac{V\ln(V/\Ub)}{\Ub}\frac{\fkz}{f_{k_0+1}}\,.
\end{equation}
According to our continuous approximation, $\ln(f_{k_0+1}/\fkz)=\partial \ln 
\fkz/\partial k_0 = \phi'(x_0).$ Thus, with $\phi'(x_0)=\ln(V/\Ub)$ 
[Eq.~\eqref{eq:u-lnu-AE}], we find
\begin{equation}
 \frac{|S|\fkz}{M}=\ln\frac{V}{\Ub}\gg 1\,.
\end{equation}
Therefore,  we can safely neglect the mutation term $M$ in the log-derivative of $\fkz$ in time and approximate it with $|S|$, which fact will be used in the derivation below. 

The dynamics of the current least-loaded class $\fkz(t)$ above the stochastic threshold $1/(N|S|)$ represents a periodic saw-shaped dependence (Fig. 6). When beneficial mutations occurring in the least-loaded class generate a new least-loaded class with a size on the order of the stochastic threshold, with probability on the order of  $1$, the new class will survive random drift and be further amplified by selection. We refer to this event as "a class is established". After a class is established, the wave moves one notch in $k$. By definition, the time period between establishment of two consecutive classes is given by $t_{\rm period}=1/V$. Below we estimate the size of the best-fit class $f_{\rm max}\equiv \fkz(t_{\rm period})$ at the moment it gives rise to a new established class.
 
\begin{figure}
\centerline{\includegraphics[width=4in]{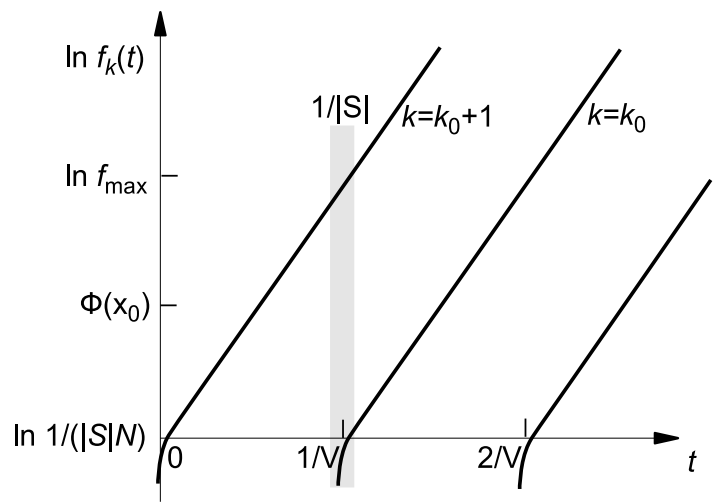}}
\caption{\label{fig:gain-of-new-class} 
Schematic illustration of establishment of a new best-fit class.
Once a beneficial mutant has survived drift, 
its frequency grows approximately exponentially in time. A further beneficial 
mutation is likely to arise and survive drift only in a short time interval of 
length $1/S$ around the time when the currently least-loaded class has grown to 
its maximal value $f_{\max}$.}
\end{figure}

The total number 
of mutational opportunities (number of genomes that may potentially generate a beneficial mutation during one click) is $N$ multiplied by the time integral of
$\fkz(t)$ over one period, which is $Nf_{\rm max}/|S|$. In other words, 
because of the exponential growth of $\fkz(t)$, most of the mutational 
opportunities will arise in a short time interval of length $1/|S|$ around the 
time when $\fkz(t)$ has come close to the value $f_{\rm max}$ (Fig.~\ref{fig:gain-of-new-class}).
We note that that time interval is much shorter than the time of one click, $1/|S|=(1/V)/\ln(V/\Ub)\ll 1/V$. We obtain the average number of new least-loaded genomes created during one period  by multiplying the number of mutational 
opportunities by the beneficial mutation rate $\Ub$. Finally, each of these 
beneficial mutations have a probability $2|S|$  to survive drift 
\citep{Haldane27,Kimura65}. Therefore, the total number of beneficial mutations 
that arise and go to fixation during one period is $2|S|\Ub 
Nf_{\rm max}/|S|=2\Ub Nf_{\rm max}$.  We find the desired value of $f_{\rm max}$ from the condition that, on the average, one new fitness class is established, which yields
\begin{equation}
\label{eq:def-fmax}
  f_{\rm max} \approx \frac{1}{2N\Ub}\,.
\end{equation}

Based on Fig. 6 and Eq.\eqref{eq:def-fmax}, we can also estimate  the time of one click $1/V$ in terms of
$S=sx_0$, as given by
 \begin{equation}
\label{tauedge}
\frac{1}{V}=  \ln f_{\rm max}-\ln\frac{1}{N|S|}=\frac{1}{|S|}\ln\frac{|S|}{U_b}
\end{equation}
which is similar to the expression $1/V=(1/|S|)\ln[V/(eU_b)]$ obtained from the deterministic part of our derivation, Eq.~\eqref{eq:x-zero-AE}. The difference is in the logarithmic factor $|S|/V\sim\ln(V/U_b)$ in the argument of the large logarithm in Eq.~\eqref{tauedge}. At the moment, we do not understand the discrepancy. The fact that  $V$ as a function of $|x_0|$ can be obtained from considering either the stochastic edge \citep{DesaiFisher2007, Brunetetal2007} or the deterministic "bulk" is certainly not trivial and deserves further inquiry.

The above estimate of $f_{\rm max}$ operates with fitness classes and does not depend on the particular clone composition of a class. Still, for the sake of clarity, we would like to make a comment on the clone composition of the least-loaded class. The estimate of $f_{\rm max}$ (defined within the accuracy of a numeric factor on the order of 1) corresponds to the short interval of time $\sim 1/|S|$ when a new least-loaded class is in the vicinity of the stochastic threshold, $\fkz\sim 1/(N|S|)$ \citep{KimuraOhta73,Barton95,Rouzineetal2001}. At that time, the new class is comprised of a small number of clones (most likely, one or two). Later, as the next-loaded class increases nearly exponentially in time above $f_{\rm max}$, it generates additional clones joining the least-loaded class. Because the additional clones cross the stochastic threshold later than the first clone(s), they are consecutively smaller in size.  As a result, the growing class consists of an increasing number of clones with decreasing relative sizes. However, the eventual clone composition of the best-fit class does not influence the estimate of $f_{\rm max}$, because it is formed at later times.

Now, following our general approach, Eq.~\eqref{match}, we 
match $\phi(x_0)$ to the average between the minimum and the maximum value of 
$f_k(t)$ under deterministic growth (Fig.~\ref{fig:gain-of-new-class}),
\begin{align}\label{eq:phi-xzero-new-AE}
  \phi(x_0) &= \frac{1}{2}\Big(\ln f_{\max} + \ln\frac{1}{|S|N}\Big) \notag \\
   & \approx -\ln\Big[N\sqrt{ V\Ub \ln(V/\Ub)}\Big]\,.
\end{align}
In comparison to the previous result by \citet{ Rouzineetal2003} who matched $\phi(x_0)$ directly to the stochastic threshold, the result found 
with the improved treatment of the stochastic edge differs by a factor of 
$\sqrt{(V/\Ub)/\ln^3(V/\Ub)}$ multiplying $N$.

As in the case of Muller's ratchet, we can calculate a correction to the 
continuous approach that accounts for the deterministic 
perturbation of the wave shape caused by the discreteness of fitness class near 
$k=k_0$. The calculation is similar to the one for Muller's ratchet, and the 
details are given in the appendix. The final expression for this correction term, 
Eq.~\eqref{eq:discont-corr-adaptation}, reads
\begin{equation}
\label{discretecorradapt}
\ln\left[2(k-k_0)+1+{\cal O}\Big(\frac{1}{\ln(V/\Ub)}\Big)\right]\,.
\end{equation}
We now replace $k-k_0$ by $|x_0|$, and neglect the term $+1$ in the above 
expression and the term $-Vs$ in $|x_0|$ [Eq.~\eqref{eq:x-zero-AE}], using the strong 
inequality $\ln(V/\Ub)\gg1$. Then, the correction term becomes
\begin{equation}\label{eq:discont-corr-adaptation-textbody}
\ln\left(\frac{V}{s}\ln\frac{V}{\Ub}\right)\,.
\end{equation}

Inserting Eqs.~\eqref{eq:phi-xzero-new-AE} and~\eqref{eq:phi-zero-AE} into 
Eq.~\eqref{eq:phi-zero-phi-x-zero-AE}, adding the correction 
term~\eqref{eq:discont-corr-adaptation-textbody} to the right-hand side of 
Eq.~\eqref{eq:phi-zero-phi-x-zero-AE}, and dropping all numerical constants 
multiplying $N$ inside the large logarithm, we arrive at our final result for 
large $V$, 
\begin{equation}\label{eq:finalspeed1}
 \ln N \approx \frac{V}{2s}\Big(\ln^2\frac{V}{e\Ub}+1\Big)- 
\ln\sqrt{\frac{s^3\Ub}{V^2\ln(V/\Ub)}}\,,
\quad V\gg s\,.
\end{equation}
For intermediate $V$, $\phi(0)=0$  from Eq.~\eqref{eq:phi-zero-AE-narrow}, and the 
final result reads 
\begin{equation}\label{eq:finalspeed2}
 \ln N \approx \frac{V}{2s}\Big(\ln^2\frac{V}{e\Ub}+1\Big)- 
\ln\sqrt{\frac{s^2\Ub}{V \ln(V/\Ub)}}\,,
\quad s/\ln(V/\Ub)\ll V\ll s\,.
\end{equation}
The difference between these two results is a factor of $\sqrt{V/s}$ multiplying 
the large population size $N$. As we found in simulations, the numeric effect of that difference is quite modest in a very broad parameter range. As in the case of Muller's ratchet, we can evaluate expressions \eqref{eq:finalspeed1} and \eqref{eq:finalspeed2} either by 
calculating $N$ as a function of $V$, or by numerically solving for $V$ at a given 
value of $N$.

As in the case of Muller's ratchet, we can express Eqs.~\eqref{eq:finalspeed1} 
and~\eqref{eq:finalspeed2} in terms of variables encountered in the standard 
diffusion limit, $V'=NV$, $s'=Ns$, and $\Ub'=N\Ub$. First, note that $V/s=V'/s'$ 
and $V/\Ub=V'/\Ub'$. Second, after subtracting $\ln N$ from both sides of 
Eqs.~\eqref{eq:finalspeed1} and~\eqref{eq:finalspeed2}, we obtain exactly the 
diffusion limit scaling for the terms $\ln\sqrt{s^3\Ub/V^2}$ and 
$\ln\sqrt{s^2\Ub/V}$ on the right-hand sides of these two equations.

For extremely large $N$, we can obtain an explicit expression for $V$ from 
Eq.~\eqref{eq:finalspeed1} through iteration. In zero approximation, we have 
$V\sim s$. We insert this value into the logarithms in Eq.~\eqref{eq:finalspeed1}, substitute the 
resulting expression again into Eq.~\eqref{eq:finalspeed1}, and then neglect all 
numeric constants and terms of the form $\ln(s/\Ub)$ inside of large logarithms. 
We also neglect $\ln N$ when it multiplies $N$ in the argument of a logarithm 
(because $N$ is assumed to be very large). We find, to the first order,
\begin{equation}\label{eq:finalspeed-huge-N}
  V \approx \frac{2s\ln( N\sqrt{s\Ub})}{\ln^2[(s/\Ub)\ln( N\sqrt{s\Ub})]}\,.
\end{equation}
The main difference from the previous asymptotic result of 
\citet{Rouzineetal2003} is the factor $\sqrt{sU_b}$ instead of $U_b$ multiplying 
$N$. [Note that in the corresponding equation in \citep{Rouzineetal2003}, Eq.~(33) 
in the Supplementary Text, the factor $\Ub$ multiplying $N$ was accidentally 
omitted.]

We assumed that only the best-fit class should be treated stochastically and that the next-loaded class, $k=k_0-1$, can 
be treated deterministically. The validity condition can be found from the requirement that the average size of the next-best class is much larger than the size of the best-fit class, which is either on the order of or larger than then stochastic threshold. Indeed, in this case the random drift term in the diffusion equation for the next-loaded class can be neglected as compared to the selection term. The requirement also ensures that, when a best-fit class gives rise to a new best-fit class, the former is much higher than the stochastic threshold. The validity condition is $ V\gg U_b$, which is equivalent to $s|x_0| \gg U_b$ (Simplification 7). Because we already assumed the strong inequality $V\gg U_b$ when deriving Eqs.~\eqref{eq:finalspeed1}, \eqref{eq:finalspeed2}, and \eqref{eq:finalspeed-huge-N}, the next-loaded class can safely be considered deterministic in the entire 
region where the results of this section apply. Thus, unlike in the case of Muller's 
ratchet, where the approximation of a single stochastic class introduces an error corresponding to a numeric factor multiplying $N$, in the case of adaption, the approximation is asymptotically accurate.

\subsection{Comparison with simulation results for the adaptation rate}

\begin{figure}
\centerline{\includegraphics[width=2.5in]{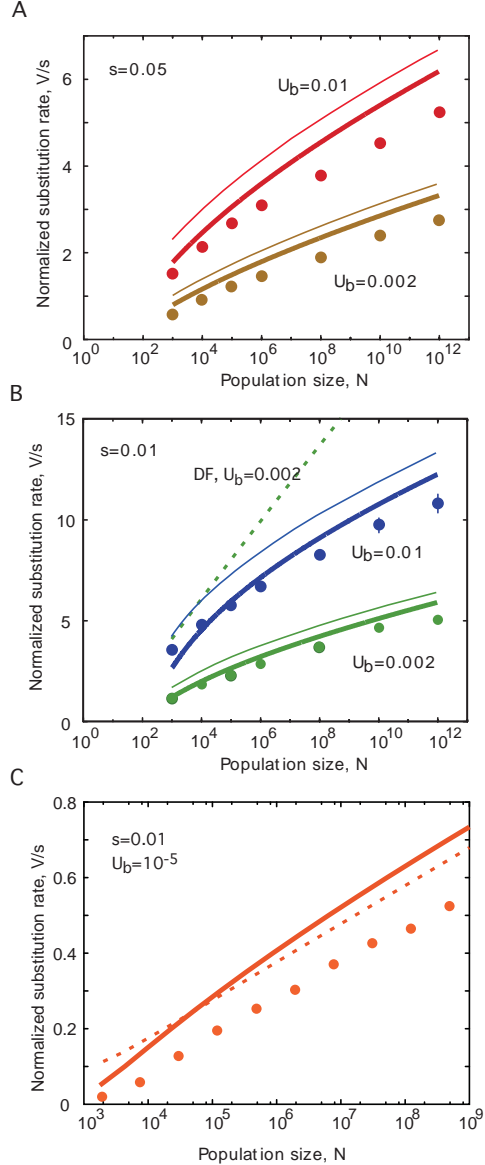}}
\caption{\label{fig:adaptation}
Normalized substitution rate as a function of population size, in the absence of 
deleterious mutations. Symbols correspond to simulation results, and 
thick solid lines correspond to analytic predictions obtained (A,B)  from 
Eq.~\eqref{eq:finalspeed1} or (C) from Eq.~\eqref{eq:finalspeed2}. Thin lines in (A,B) indicate the analytic prediction without discreteness correction, i.e., without term $\ln[(V/s)\ln(V/\Ub)]$ on the 
right-hand side of Eq.~\eqref{eq:finalspeed1}. Dotted lines in (B,C) indicate the analytic result by 
\citet{DesaiFisher2007}, Eqs. 36, 38, and 39. Their result shown in (B) is outside of its designed validity range and is shown for reference only. The simulations were carried out (A,B) in discrete time as 
described \citep{Rouzineetal2003} or (C) in continuous time. $V$ was measured in simulation as the average slope of $\kav(t)$ in the time interval $[0.85t_0, 1.15t_0]$, where the 
time $t_0$ of the interval center was determined from the condition 
$\kav(t_0)=250$. The initial $\kav(0)$ was set to $1.5\kav(t_0)$. The per-site 
beneficial mutation rate $\mu$ was chosen such that the genomic beneficial 
mutation rate $\Ub$ had the value shown on the plot at time $t_0$, 
$\Ub=\mu\kav(t_0)$. The size of the symbols roughly corresponds to the largest standard 
deviation of the mean speed estimates.
}
\end{figure}

We carried out simulations of adaptive evolution as described 
\citep{Rouzineetal2003}, and compared the measured speed of adaptation to the 
speed predicted by Eq.~\eqref{eq:finalspeed1} in a wide range of parameter settings 
(Fig.~\ref{fig:adaptation}A-C). Because deleterious mutations can be neglected in the limit $V\gg U$ we consider, the simulation results we 
present here were obtained in the absence of deleterious mutations, $U=\Ub$, 
whereas \citet{Rouzineetal2003} considered beneficial and deleterious mutations at 
the same time.

Without the discreteness correction, Eq.~\eqref{discretecorradapt}, the analytic prediction of the wave speed generally follows the simulation results fairly well in a broad parameter range. The analytic result consistently overestimates the simulation result  by 15 to 25\% (Fig.~\ref{fig:adaptation}A to C, dashed line). When we include the discreteness 
correction, the accuracy improves significantly across all parameter settings. In general, the accuracy becomes higher as $\Ub$ and $s$ decrease. (If deleterious mutations are present, $U>\Ub$, the predictions from traveling wave 
theory are at least as accurate---if not more so---as in the absence of 
deleterious mutations; see Fig.~2f in \citet{Rouzineetal2003} ). 

For comparison, we show analytic results for the substitution rate obtained by \citet{DesaiFisher2007} who used a different approach [Eqs. (36), (38), and (39) in the cited work]. Their method applies at small adaptation rates where the lead $|x_0|$ is only moderately large (see our \emph{Discussion}). In its validity range, the prediction is very similar to ours (Fig.~\ref{fig:adaptation}C, dotted line). Note that \citet{DesaiFisher2007} in their Fig. 5 compare with simulation not the actual analytic result, but its simplified version, Eq. (41), which accidentally gives a higher accuracy than the actual result. When the lead is large and the approach breaks down, their analytic prediction is far from simulation results (Fig.~\ref{fig:adaptation}B, dotted line). 

\subsection{Testing stochastic edge treatment: Simulation of a two-class model}

In order to validate our results and to get further insight into the dynamics of the stochastic best-fit class, we carried out a computer simulation for a simplified model including only two fitness classes, the best-fit class, and the second-best class \citep{DesaiFisher2007} .  Only the best-fit class was treated stochastically. The aim was to confirm the analytic estimate of the size of the second-best class when the new best class is established, Eq.~\eqref{eq:def-fmax}. The way it was derived, the estimate is expected to be robust with respect to such a model simplification. Note that the two-class model cannot be used to test other intermediate results of this section.

According to the two-class model, we assume that the frequency of the second-best class grows as
\begin{equation}\label{eq:fk0-1-of-t}
  f_{k_0-1}(t) = \frac{1}{Ns|x_0|}e^{s(|x_0|-1)t},
\end{equation}
where $1/(s|x_0|)$ is the characteristic size where selection prevails over genetic drift. (In other words, we ignore stochastic effects  and mutants coming from less-fit classes.) According to the main model, the {\it average} frequency of the best class grows as
\begin{equation}\label{eq:n-of-t}
  f_{k_0}(t+1) = e^{s|x_0|}  f_{k_0}(t) + \Ub f_{k_0-1}(t)= e^{s|x_0|} f_{k_0}(t) + \frac{\Ub}{Ns|x_0|}e^{s(|x_0|-1)t}.
\end{equation}
In our pseudorandom simulation, we calculated  the actual size of the best class $Nf_{k_0}(t)$ over a long time scale using Poisson statistics with the average given by Eq.~\eqref{eq:n-of-t}. Then, to find the time delay between the growth of two classes, we extrapolated $f_{k_0}(t)$ back from large times $t\gg 1/s|x_0|$ when the best-fit  class was large enough to be treated deterministically,  $Nf_{k_0}\gg 1/s|x_0|$, as follows.

We seek a solution of Eq.~\eqref{eq:n-of-t} in the form $f_{k_0}(t)=A(t)e^{s|x_0|t}$. Inserting this expression into 
Eq.~\eqref{eq:n-of-t} and solving the equation in discrete time with respect to $A(t)$, we obtain
\begin{equation}\label{fk0-vs-A0}
  f_{k_0}(t)= e^{s|x_0|t}\left[A(0) + \frac{\Ub e^{-s|x_0|}}{s|x_0|}\frac{1-e^{-st}}{1-e^{-s}}\right],
\end{equation}
where $A(0)$ is an arbitrary constant. Strictly speaking, $A(0)$ is random and has to be found from the stochastic edge simulation. 

Instead of $A(0)$, we introduce a more transparent parameter, the time delay between establishment of two consecutive best-fit classes $\tau$, as given by the periodicity condition 
\begin{equation}\label{tau-periodic}
  f_{k_0}(\tau)=f_{k_0-1}(0)=\frac{1}{Ns|x_0|}.
  \end{equation}
 From Eq.~\eqref{fk0-vs-A0}, the parameter $\tau$ is related to $A(0)$ as given by 
\begin{equation}
  A(0) = \frac{1}{s|x_0|}\left[{e^{-s|x_0|\tau}+\Ub e^{-s|x_0|}\frac{e^{-s\tau}-1}{1-e^{-s}}}\right].
\end{equation}
In terms of $\tau$, the growth curve of the best-fit class,  Eq.~\eqref{fk0-vs-A0}, can be written as
\begin{equation}\label{eq:final1}
 f_{k_0}(t)= \frac{1}{Ns|x_0|}e^{s|x_0|t} \left[ {e^{-s|x_0|\tau}+\Ub e^{-s|x_0|}\frac{e^{-s\tau}-e^{-st}}{1-e^{-s}}}\right].
\end{equation}
We obtain $\tau$ by measuring $f_{k_0}(t)$ in simulation at some large time $t$, such that $t\gg 1/s|x_0|$, and then solving Eq.~\eqref{eq:final1} numerically for $\tau$. We checked that changing the sampling time $t$ within the range $\tau$ to $10\tau$ has a small effect on the result for $\tau$, which confirms the vadity of our back-extrapolation procedure. 

Finally, in order to test the accuracy of our analytic estimate, Eq.~\eqref{eq:def-fmax}, we calculate the constant $C$ in the stochastic edge simulation, as given by
\begin{equation}\label{C}
  C = N\Ub f_{k_0-1}(\tau)\equiv  N\Ub f_{\rm max}.
\end{equation}

\begin{figure}
\centerline{\includegraphics[width=4in]{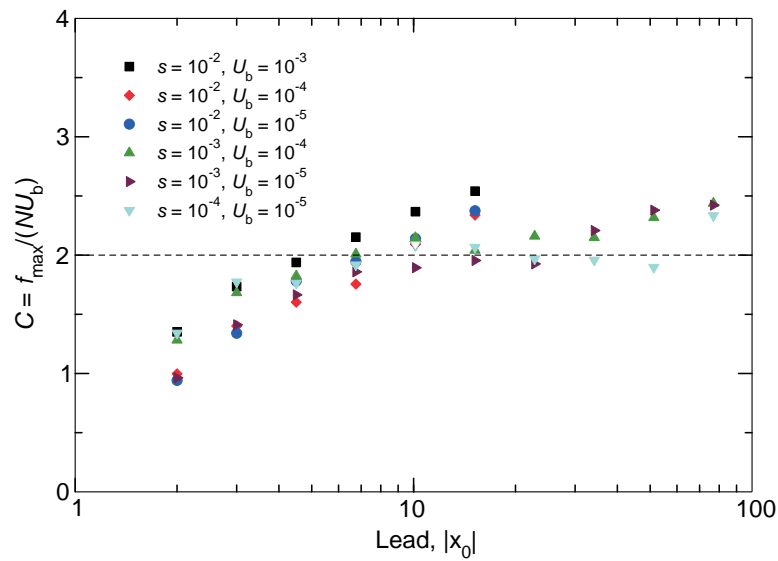}}
\caption{\label{fig:testC}
Test of the accuracy of the stochastic edge treatment,  Eq.~\eqref{eq:def-fmax}. Points are obtained by simulation using a simplified two-class model. Only points corresponding to $s|x_0| < 0.2$ are shown. Parameter values are shown. }
\end{figure}

We averaged the value of $\log C$ over 500 simulation runs. The result is shown in Fig.~\ref{fig:testC} for various values of $|x_0|, s$, and $U_b$. In principle, the analytic result that $C=\text{const}\approx 2$ is supposed to be asymptotically correct at  $|x_0|\gg1, s|x_0|\ll 1$. In the parameter region $|x_0| > 5$ and $s|x_0| < 0.2$, the value of $C$ varies between 1.5 and 2.5 . (Note that even though Eq.~\eqref{C} seems to imply that $C$ 
depends on $N$, this factor actually cancels.) Because $C$ is only a numeric factor at  $N$, and $N$ is large and enters a logarithm in the expression for the substitution rate, such variation of $C$ has a small effect on the final substitution rate  and thus  confirms the validity of our analytic estimate,  Eq.~\eqref{eq:def-fmax}.

It is important that, due to constant influx of  new beneficial mutants from the second-best class, the best-fit class size does not grow exactly exponentially,  Eq.~\eqref{eq:final1}. We checked that replacing Eq.~\eqref{eq:final1} with an exact exponential in the back-extrapolation procedure seriously affects the result for $\tau$ and makes it depend on the (arbitrarily chosen) sampling time $t$ (results not shown). This facts explains why the approach by \citet{DesaiFisher2007}, who used the exponential  back-extrapolation to estimate $\tau$, works only at very small adaptation rates when $|x_0| \sim 1$  [Fig. ~\ref{fig:adaptation}) (see \emph{Discussion} and \citet{Brunetetal2007}].

\subsection{Crossing over to the single site model}

We expect intuitively that, for extremely large population sizes, the speed of 
adaptation should cross over to its deterministic, infinite-population-size 
behavior. In the deterministic limit, the process of creation and fixation of rare 
mutants does not affect the speed of adaptation, because all possible mutants are 
instantaneously present. Instead, the speed of adaptation depends simply on the 
growth rate of the mutants at different sites. As a consequence, in the 
mulitplicative fitness model, linkage between sites becomes irrelevant, and we can 
describe the system behavior with the single site model,
\begin{equation}
	\frac{df}{dt} = sf(1-f) +\mub(1-f)\,,
\end{equation}
where $f$ is the frequency of the beneficial allele at the given site, and we have 
assumed that deleterious mutations do not occur. The time it takes for a 
beneficial clone to grow from size 0 to size $f_0$ is $T=\ln[(sf_0 + \mub)/(\mub-
f_0\mub)]/(s+\mub)$. Setting $f_0=1/2$ and assuming $s\gg\mub$, we find that the 
half-time of reversion, in which the beneficial clone grows to 50\% presence, is
\begin{equation}\label{half-time-single-site}
  T_{1/2} \approx (1/s)\ln(s/\mub)\,.
\end{equation}
The same equation applies to a population of genomes that contain multiple sites 
which are independent and unlinked. Thus, if at some time point there are $\kav$ 
deleterious alleles fixed in the population, then Eq.~
\eqref{half-time-single-site} gives the time interval until, on average, all members of the population 
have reverted half of these deleterious alleles.

To test our intuitive expectation, we will now compare this expression to the 
corresponding expression from traveling wave theory. For a wave centered at 
position $\kav$, the time to revert half of the deleterious mutations is
\begin{equation}\label{half-time-general}
  T_{1/2} \approx \frac{\kav}{2V}\,.
\end{equation}
After inserting Eq.~\eqref{eq:finalspeed-huge-N} into Eq.
~\eqref{half-time-general} and keeping only the leading terms in $N$, we find that at $\ln N\sim 
\kav\ln(s/\mub)$, the half-time of reversion crosses over to the single-site 
expression Eq.~\eqref{half-time-single-site} (under the assumption that 
$s\gg\mub$). The reason for this behavior is that once $\ln N$ grows to 
$\kav\ln(s/\mub)$, the left edge of the wave reaches the mutation-free genome. At 
this point, the assumption $\kav \gg \kav-k_0$ is violated, and the traveling-wave 
approach ceases to be valid. For population sizes exceeding $\kav\ln(s/\mub)$, the 
system behavior is essentially independent of population size, and linkage does 
not slow down the speed of adaptation.

\section{Discussion}

We have derived accurate expressions for the speed of Muller's ratchet and the 
speed of adaptation in a finite, asexual population. Our expressions are numerically valid in 
a wide parameter range. The main difference between 
the work presented here and the work of \citet{Rouzineetal2003} is an improved and 
more intuitive treatment of the stochastic egde, and the analytic derivation of 
correction terms for the discontiuity at $k=k_0$ for both Muller's ratchet and the 
speed of adaptation. The latter correction terms lead to significantly improved 
numerical accuracy in a wide range of population sizes.

\subsection{Muller's ratchet}
A quantitative treatment of Muller's ratchet was first attempted by 
\citet{Haigh78}, who studied primarily the case in which ratchet clicks are rare, 
so that the population can equilibrate between subsequent ratchet clicks. Since 
then, numerous studies have derived improved estimates of the ratchet rate 
\citep{Pamiloetal87,Stephanetal93,Gessler95,HiggsWoodcock95,PruegelBennett97,
Lande98,GordoCharlesworth2000a,GordoCharlesworth2000b}. Most approaches to calculate 
the ratchet rate work either in the case $N<1/s$, when fixation events of 
deleterious alleles at different sites are well separated in time and the one-site 
theory applies, or in the limit of very large population sizes, when the 
deterministic expectation for the size of the fittest class $n_0=N\exp(-U/s)$ is 
much larger than 1. In the former case, the ratchet rate is estimated as the rate 
at which deleterious mutations enter the population, multiplied with the 
probability of fixation of deleterious mutations \citep{Lande98}. In the latter 
case, the ratchet rate is estimated from the average time the fittest class takes 
to drift from its equilibrium value to an occupancy of zero  
\citep{Stephanetal93,GordoCharlesworth2000a,GordoCharlesworth2000b}. 
A third approach, most closely related to the present work, is a quantitative 
genetics approach whereby the fitness distribution is described by its mean, 
variance, and higher moments, and equations are derived that describe the change 
of these moments over time \citep{Pamiloetal87,HiggsWoodcock95,PruegelBennett97}. 
A disadvantage of this approach is that it generally requires an arbitrary cutoff 
condition to truncate the moment expansion, and that the equations become quickly 
untractable if higher moments are included.

The rapid ratchet region described in the present work corresponds, mostly, to 
$n_0$ smaller than 1 (Fig. 5c). For small $n_0$, the work by  \citet{Gessler95} is 
generally regarded as giving accurate estimates for the ratchet rate. However, it 
is important to emphasize that \citet{Gessler95} did not actually derive a closed 
form expression for the ratchet rate. A key parameter in his result needs to be 
determined numerically by iteration. Moreover, several of his expressions were not 
derived from first principles, but chosen on the basis that they provided a good 
fit to simulation results. In contrast, our expression Eq.~\eqref{finalratchet} 
was derived from first principles and is a simple, closed-form expression that can 
be plotted easily, if we interpret it as describing $N$ as a function of $v$ 
(rather than $v$ as a function of $N$).

The traveling wave approach also provides a convenient framework to study the 
ratchet speed in the presence of beneficial mutations, or, conversely, the speed 
of adaptation in the presence of deleterious mutations. All that is required is to 
keep the full expression $\phi(0)-\phi(x_0)$, 
Eq.~\eqref{eq:delta-phi3}, instead of its respective asymptotics. While we did not discuss these cases here in detail, 
the corresponding calculations are straightforward and lead to closed-form 
expressions [see \citep{Rouzineetal2003}, Supplementary Text]. In comparison, 
other approaches may lead to similarly accurate expressions, but usually do not 
yield closed-form expressions. For example, \citet{BachtrogGordo2004} studied the 
effect of beneficial mutations in the presence of Muller's ratchet. They used a 
Poisson approximation for the steady-state distributions of mutations in 
populations undergoing Muller's ratchet, explicitly modeled the behavior of 
beneficial mutations arising in all possible mutation classes, and used the 
resulting fixation rates as corrections for the ratchet rate derived by 
\citet{Gessler95} for $n_0<1$  and by 
\citet{GordoCharlesworth2000a,GordoCharlesworth2000b} for $n_0>1$. This approach 
leads to accurate results in a wide parameter range, but the final equations 
contain sums over all mutation classes, and these sums cannot be evaluated 
analytically.

Finally, with the traveling wave approach, it is also possible to consider the 
general case when both beneficial and deleterious mutations are equally important, 
including steady state at finite $N$, $v=0$ in Eq.~\eqref{eq:delta-phi3}, which 
differs from equilibrium at infinite $N$ \citep{Rouzineetal2003}. However, to 
achieve a high accuracy in this case requires some additional work on the 
stochastic edge condition.

\subsection{Adaptation}
In an asexual population, the speed of adaptation does not grow linearly with $N$ 
for large $N$, because beneficial mutations that arise contemporaneously in 
independent genetic backgrounds cannot recombine, and, therefore, only a small 
fraction of all beneficial mutations that survive drift can actually go to 
fixation \citep{Fisher30,Muller32,CrowKimura65,HillRobertson66,MaynardSmith71}. 
One reason why a beneficial mutation may not go to fixation is interference with 
another mutation with a larger beneficial effect that arises either shortly before 
or shortly after the first mutation \citep{Barton95}. Recently, a theory that 
predicts the evolutionary rate in large asexual populations in the presence of 
this interference process has received considerable attention (clonal 
interference-theory, \citealp{GerrishLenski98,Orr2000,Wilke2004}). However, 
interference does not necessarily occur in the presence of a single alternative 
mutation. In fact, multiple beneficial mutations arising in the already existing 
clones --- that have moderate individual effect but large combined effect---can 
rescue clones with a smaller beneficial effect and partly resolve the clonal 
interference. The clonal interference theory neglects this effect and predicts 
that the substitution rate rapidly approaches a constant with increasing $N$  
(under the assumption of exponentially distributed beneficial effects, 
\citealp{Wilke2004}), and that further increases in the speed of adpatation can 
only come from fixation of mutations with increasingly larger beneficial effects. 
This prediction is, most likely, not entirely correct, because an increase in 
population size also increases the chance that multiple mutations occur in the 
same clone. In general, we expect the clonal interference theory to have a very 
limited range of applicability. If beneficial mutations are rare, interference is 
irrelevant to the speed of adaptation, and $V$ is proportional to $N\Ub$. On the 
other hand, if the population size is large and beneficial mutations are frequent, 
then multiple mutations within a single clone should be frequent as well, and we 
expect the current clonal interference theory to underestimate the 
speed of adaptation in this regime.

In contrast to the clonal interference approach, the traveling wave approach 
considers multiple mutations. However, the present model does not consider 
interference from mutations with a larger beneficial effect, because all mutations 
are assumed to have the same selection coefficient. (Note that, as a consequence, the 
substitution rate and the speed of adaptation are essentially equivalent in our 
model, as they differ only by a constant factor $s$, whereas they are distinct 
quantities if there is variation in the selection coefficient, as is the case in 
the clonal interference models.) Because of these differences in model 
assumptions, we cannot make a quantitative comparison between the two theories. 
However, we can observe an important qualitative difference: in the 
traveling-wave theory, the substitution rate keeps slowly increasing with population size 
until extremely large population sizes.

For very large population sizes, such that $\ln 
N\sim\kav\ln(s/\mub)$, the substitution rate becomes on the order of $sk_{\rm av}$. At this point, the tail of the wave reaches the best possible genome, $k=0$, and the traveling wave theory breaks down \citep{Rouzineetal2003}. At higher population sizes, the one-locus model applies, and the value of the substitution rate saturates at the infinite-size value. Thus, we predict a continuous transition from the finite-population case in which linkage slows down the speed of adaptation to the infinite-population case in which linkage has no effect on the speed of adaptation in the multiplicative model 
\citep{MaynardSmith68}. By contrast, \citet{MaynardSmith71} argued that a finite 
asexual population is always affected by linkage, regardless of its size. We must 
mention, however, that, for reasonable parameter settings, the population size at 
which the crossing over to the fully deterministic behavior occurs is unbiologically large. For example, assuming $\kav=20$ and $s/\mub=1000$, we find that $N$ has to be on the order of $10^{60}$, 
and this number grows rapidly for larger $\kav$.

It is important to stress that the speed of adaptation is, in general, not 
governed by the total supply of beneficial mutations, $N\Ub$, because $N$ and 
$\Ub$ enter the expressions for the speed of adaptation independently and in 
different functional forms. This fact had already been noted by 
\citet{MaynardSmith71}, and its implications for experimental tests of the speed 
of adapation have been discussed recently in the context of clonal interference 
theory \citep{Wilke2004}.

Recently, \citet{DesaiFisher2007} have proposed an alternative method to derive, for the same initial model, the speed of adaptation in a large asexual population in the absence of deleterious mutations. We discuss this method in detail and extend its validity range elsewhere \citep{Brunetetal2007}. Briefly, these authors replace the full population model with a simplified model including only two fitness classes, the least-loaded class and the next-loaded class. The next-loaded class is assumed to grow exactly exponentially, as if under selection alone. The click time $1/V$ is determined by extrapolating the growth of the least-loaded class back from infinite time. As in our work, only the least-loaded class is treated stochastically. 

The main validity conditions for this approach are, as follows. The next-loaded class can be treated deterministically if $s|x_0|\gg U_b$, which is equivalent to $V\gg  U_b$, the same condition that we use in our work. (Note that inequality $s\gg U_b$ given by \citet{DesaiFisher2007} was obtained for  $|x_0|\sim 1$; for general $|x_0|$, the condition is as we state.) The back-extrapolation procedure is valid at moderate population sizes or adaptation rates, such that $\ln|x_0|\ll \ln(s/U_b)$  \citep{Brunetetal2007}. A sufficient validity condition is $|x_0|\ll \ln(s/U_b)$, which is equivalent to $V\ll s$ (note that the cited paper contains a typo in that condition; M. Desai, pers. comm.). The accuracy of the replacement of the full population model with the two-class model and of the actual dynamics of the next-loaded class with an exponential is less clear. New beneficial mutations into the class create a time-dependent prefactor at the exponential. Yet, it is quite possible, that at moderate $|x_0|$ or $V$ where the cited approach is designed to apply, the approximation works well. We tested the numeric accuracy of the substitution rate predicted by \citet{DesaiFisher2007} with the use of Monte-Carlo simulation of the full population model in different parameter regions. Within its validity  range,  their result agrees with simulation within 10-20\% and is very similar to our result (Fig. 7C). [Note that the extremely high numeric accuracy shown in Fig. 5 of the cited work is due to replacement of the actual final result, Eq. (39), with its asymptotics for very large $N$, Eq. (40), which does not apply at such moderate $N$.] 

In their Discussion section, \citet{DesaiFisher2007} state that the traveling wave approach \citet{Rouzineetal2003} requires that the fitness distribution is smooth in mutational load and, hence, broad, which takes place at large substitution rates, $V\gg s$. Proposing some estimates of parameter values of $U_b$ and $s$ typical for yeast, they argue that this regime corresponds to huge population sizes and conclude that results by \citet{Rouzineetal2003} are relevant for viruses only. In fact, the traveling wave approach requires that the logarithm of the fitness distribution, not the fitness distribution itself, is smooth in the mutational load, which takes place when the high-fitness tail of the distribution $|x_0|$ (not the half-width) is larger than 1. The condition is met at much smaller substitution rates, $V\gg s/\ln(V/U_b)$, which is the low bound assumed by  \citet{DesaiFisher2007}. The likely reason for the confusion is that \citet{Rouzineetal2003} normalized the fitness distribution assuming it is broad, which corresponds to the case $V\gg s$. When  the wave is narrow, which happens in the interval $s/\ln(V/U_b)\ll V\ll s$,  the normalization condition is simply replaced with $\phi(0)=0$ (Subsection 2.4). Except for the trivial change in normalization, whether the wave is broad or narrow does not matter. In the final expression for the substitution rate, the difference between the two cases $s/\ln(V/U_b)\ll V\ll s$ and $V\gg s$ is in a constant factor at $N$,  Eqs. \eqref{eq:finalspeed2} and \eqref{eq:finalspeed1}. Numerically, the two expression are quite similar in the parameter range we test. Therefore, the results of the traveling wave approach apply in both intervals of the substitution rate and are potentially relevant for  a broad variety of asexual organisms, including RNA and DNA viruses, yeast, coral, and some species of plants and fish. 

To summarize, we have applied the semi-deterministic traveling wave approach to 
derive the rate of Muller's ratchet and the speed of adaptation in a  multi-site 
model of an asexually reproducing population. Our results underscore the 
importance of a proper stochastic treament of the best-fit class. Our method, 
based on a continuous approximation to logarithm of the fitness distribution of the other (deterministic) 
classes, ensures fair accuracy in a very broad parameter range. In addition, an 
analytic correction for the discreteness of fitness of the deterministic classes 
significantly increases the accuracy of predictions.

\section*{Acknowledgments}

This work was supported by NIH grants R01AI0639236 to I.M.R. and R01AI065960 to 
C.O.W. Two of us (I.M.R. and E.B.) are grateful to Nick Barton and Alison Etheridge, the organizers of the 
Edinburgh 2006 Workshop on Mathematical Population Genetics, for giving us the 
opportunity to meet. We thank Michael Desai for stimulating discussion and
Nick Barton and Brian Charlesworth for useful comments.

%\bibliographystyle{elsart-harv}
%\bibliography{paper}

\begin{appendix}

\section{Appendix}

\subsection{Main simplifications of the derivation}

Below we list the main simplifications employed in the derivation of the traveling wave solution and determine parameter regions in which they are asymptotically correct.  A brief summary of the validity conditions is as follows. The values of $s, U$, and $U_b$ are assumed to be much smaller than 1. The number of sites is large and the wave is far away from the two extremes of a mutation-free or a completely mutated genome, $L\gg 1, \kav\gg |x_0|, L-\kav\gg |x_0|$. Other conditions depend on the case.

For Muller's ratchet, we assume $s\ll U\ll 1$ and $N\gg1/(U\sigma^{2/3})$. These conditions ensure that the high-fitness tail is long. Also, for the ratchet to proceed rapidly and continuously rather than by exponentially rare clicks, we need $\ln(Ns^{3/2}U^{1/2})\ll U/s$ (Section 3.3).

For the adaptation regime,  to have a single stochastic class, we assume $V\gg U_b$, which is equivalent to $U_b\ll s|x_0|$. To expand fitness in the mutational load, we assume $s|x_0| \ll 1$. We also need  a long tail of the fitness distribution, $|x_0|\gg 1$. In terms of the substitution rate $V$, the last two conditions read  $s/\ln(V/U_b)\ll V\ll 1/\ln(1/U_b)$. In terms of $N$, they read  $N\gg \sqrt{s/U_b^3}/\ln(s/U_b)$ and $\ln(N\sqrt{sU_b})\ll (1/s)\ln^2(1/U_b)$. 

\emph{Simplification~1: Linear selection term.} We can expand $e^{-sx}$ as $1-sx$ 
under the condition $s|x|\ll1$. In the case of Muller's ratchet, we find from 
Eq.~\eqref{eq:x-zero-MR} that this condition is equivalent to $U\ll 1$. In the 
case of adaptive evolution, we find from Eq.~\eqref{eq:x-zero-AE} that this 
condition is equivalent to $V\ll 1/\ln(1/U_b)$, which is met numerically in all cases we consider 
(Fig.~\ref{fig:adaptation}) and analytically at $\ln(N\sqrt{sU_b})\ll (1/s)\ln^2(1/U_b)$, 
Eq.\eqref{eq:finalspeed-huge-N}.

\emph{Simplification~2: Distribution is far from the origin.} Because 
$\alpha_k$ depends only slowly (linearly) on $k$, we are allowed to replace  $\alpha_k$ by $\alpha\equiv \alpha_{\kav}$ under the 
condition that $L\gg1$, $\kav\gg |x_0|$, $L-\kav\gg |x_0|$  (i.e., the 
wave is far away from the two extremes of a mutation-free or a completely mutated 
genome, and narrow in comparison to the average mutational load).

\emph{Simplification~3: Continuity in time.} Approximating $f_k(t+1)/f_k(t)$ with 
$1+\partial \ln f_k(t)/\partial t$ is justified when $|\partial \ln 
f_k(t)/\partial t| \ll 1$. Replacing $|\partial t|$ and $\ln f_k(t)$ with 
$|dx/(vU)|$ and $\phi(x)$, respectively, we find that this condition takes the 
form $U|v\phi'(x)|\ll 1$. Because $|\phi'|$ increases with $|x|$, the far low-
fitness tail is not very important for evolution, and in the high-fitness tail 
$|\phi'(x)| < |\phi'(x_0)| $, the sufficient condition  takes a form $U|v|\ln u 
\ll 1$, where $u$ is defined in Eq.~\eqref{eq:u-expression}.

In the ratchet limit, $\alpha=0$, after substituting  $u$ from 
Eq.~\eqref{eq:u-expand-MR}, the validity condition becomes $Uv\ln(1/v)\ll 1$, which is met at any $v$, 
$0<v<1$, as long as $U \ll 1$. Note that the condition is equivalent to $S \ll 1$, 
as in Simplification 1.

In the adaptation limit, where deleterious mutations can be neglected, the 
validity condition takes a form $V\ln(V/U_b)\ll 1$, where $V\equiv -Uv$ and we 
used Eq.~\eqref{eq:u-lnu-AE} for $u$. Again, the validity condition is $|S|\ll 1$, 
which is approximately equivalent to the condition $s|x_0|\ll 1$ we used to expand 
fitness in $k$ (Simplification~1).

For sufficiently (and unrealistically) large $N$, $V$ becomes large as well, and the condition $|S| \ll 1$ may be violated. In this region, technically, we can neither expand fitness in 
$k$ nor replace discrete time with continuous time (Simplifications 1 and 3 cannot be used). Nevertheless, 
\citet{Rouzineetal2003} showed that the continuous equation for the fitness 
distribution, Eq.~\eqref{eq:final-wave-eq}, can be derived in a more general way, 
without assuming $|\partial\ln f_k(t)/\partial t| \ll 1$, nor expanding fitness in 
$k$, nor neglecting multiple mutations per genome per generation [see the 
transition from Eq.~(1) to Eq.~(11) in the Supplementary Text of the quoted work]. 
In the present work, we use these approximations to simplify our derivation.

\emph{Simplification~4: Continuity in $k$.} Replacement of 
$\ln[f_{k+1}(t)/f_k(t)]$ with the partial derivative $\partial \ln f_k(t)/\partial 
k$ is justified if $|\partial \ln f_k/\partial k| \gg |\partial^2 \ln f_k/\partial 
k^2|$. With $\partial \ln f_k/\partial k= \phi'$, we can rewrite this condition as 
$|\phi'| \,|dx/d\phi'|\gg 1$. This condition is already discussed  in the section 
on the continuity correction for Muller's ratchet, and also in the section on the 
speed of adaptation, after Eq.~\eqref{eq:phi-zero-phi-x-zero-AE}. In short, this 
simplification is justified if the left tail is long, $|x_0|\gg 1$. Let us express this condition in terms of the substitution rate and $N$.

For Muller's ratchet, based on Eq. \eqref{eq:x-zero-MR}, the condition $|x_0|\gg 1$ is met if $\sigma\ll 1$, and the ratchet rate is not to close to the maximum, $1-v\gg\sqrt{\sigma}$. The latter condition is always met, because the more limiting condition is that the wave tail is  longer than the half-width, $|x_0|\gg\sqrt{{\rm Var}[k]}$, which is equivalent to the requirement $\phi(0)-\phi(x_0)\gg 1$. In terms of $v$, that condition reads  $1-v\gg\sigma^{1/3}$, Eqs. \eqref{eq:phi-zero-MR} and \eqref{eq:x-zero-MR}. The lower bound on $N$ follows from the final relation between $N$ and $v$, Eq. \eqref{finalratchet}, which yields
$N\gg1/(U\sigma^{2/3})$. 

For adaptation, the condition of a long high-fitness tail can be found from Eq. \eqref{eq:x-zero-AE}  for $|x_0|$, as given by $V\gg s/\ln(V/U_b)\approx s/\ln(s/U_b)$. The lower bound on $N$ follows from the final result for the adaptation speed, Eq. \eqref{eq:finalspeed2}, as given by $N\gg \sqrt{s/U_b^3}/\ln(s/U_b)$.

\emph{Simplification~5: Slow change of the wave shape.} In the derivation of 
Eq.~\eqref{eq:wave-shape-equation}, we neglected any change in the wave shape over 
time, i.e., we assumed that $|\partial\phi/\partial t|\ll1$. This condition is 
satisfied when the distribution is far from the origin, Simplification~2
[see \citep{Rouzineetal2003}, Supplementary Text, Approximation 6 for details].

\emph{Simplification~6: Broad and narrow wave.} If the wave is broad,
 $\Vark\gg 1$, to determine 
$\phi(0)$, we can expand $e^{\pm\phi'(x)}$ near the wave center linearly as 
$1\pm\phi'(x)$ if $|\phi'(x)|\ll 1$. This condition is met 
for any $v$, whether positive or negative, if $\sigma \ll 1$ (i.e., $s\ll U$).  If  
$\sigma$ is not small, the condition is is met at large negative $v$ (rapid 
adaptive evolution): $|v|\gg \sigma$ or $V\gg s$. In the opposite limit of narrow wave, we have 
$\Vark\ll 1$, so that $\phi(x)$ is narrow and cannot be approximated by a Gaussian 
even near its center. In this case, we use instead $\phi(0)=1$. Together, these two limits cover the entire parameter range.

\emph{Simplification~7: Single stochastic class in the adaptation case.} Now we verify the initial approximation that the next-loaded class, $k=k_0-1$, can 
be treated deterministically. The average ratio of its size to the least-loaded 
class size is given by $f_{k_0-1}/ f_{k_0}=\exp[\phi'(x_0)]=u=V/U_b$, 
Eq.~\eqref{eq:u-lnu-AE}. Thus, the condition that only one class has to be treated deterministically is $V\gg U_b$. The strong inequality also ensures that, when a best-fit class gives rise to a new best-fit class, the former is much higher than the stochastic threshold, $f_{\rm max}\gg 1/(Ns|x_0|)$, Eqs. ~\eqref{eq:def-fmax} and \eqref{eq:x-zero-AE}.

\subsection{Connection between the width and speed of the wave---alternative 
derivation}

We derived the maximum ratchet rate ($v<1-2\alpha$) from the variance of $k$, 
Eq.~\eqref{eq:var-k}, which in turn we derived from the normal approximation to 
the shape of the wave, Eq.~\eqref{eq:phi-prime}. We can alternatively derive the 
same result directly from Eq.~\eqref{eq:det-expanded-Ualpha}. We write $f_k(t+1)-
f_k(t) = U\partial f_k(t)/\partial \tau$, multiply both sides of 
Eq.~\eqref{eq:det-expanded-Ualpha} with $(k-\kav)/U$, and sum over all $k$. We 
find
\begin{align}
\label{eq60}
  \sum_k (k-\kav) \frac{\partial f_k(t)}{\partial \tau} &= (1-\alpha) \sum_k (k-
\kav) f_{k-1}(t) + \alpha \sum_k(k-\kav)f_{k+1}(t)\notag\\
	&\qquad\qquad - \sum_k(k-\kav)[1+\sigma(k-\kav)] f_{k}(t)\,.
\end{align}
The term on the left hand side of Eq. \eqref{eq60} evaluates to
\begin{align}
  \sum_k (k-\kav) \frac{\partial f_k(t)}{\partial \tau} &=
      \frac{\partial}{\partial \tau} \sum_k kf_k(t) - \kav 
\frac{\partial}{\partial \tau} \sum_k f_k(t) \notag\\
	& = \frac{\partial\kav}{\partial \tau}=v\,.
\end{align}
The first term on the right-hand side of Eq. \eqref{eq60} evaluates to
\begin{align}
  (1-\alpha) \sum_k (k-\kav) f_{k-1}(t) &= (1-\alpha)\Big[
      \sum_k (k-1)f_{k-1}(t) + (1-\kav)\sum_k f_{k-1}(t)\Big]\notag\\
   &= (1-\alpha)\,. 
\end{align}
Likewise,
\begin{equation}
  \alpha \sum_k(k-\kav)f_{k+1}(t) = -\alpha
\end{equation}
and
\begin{equation}
 \sum_k(k-\kav)[1+\sigma(k-\kav)] f_{k}(t) = \sigma\sum_k (k-\kav)^2 f_k(t)\,.
\end{equation}
Since $\sum_k (k-\kav)^2 f_k(t)=\Vark$, we find
\begin{equation}
  v = 1-2\alpha -\sigma \Vark\,,
\end{equation}
which corresponds to Eq.~\eqref{eq:var-k}.

\subsection{Correction to the continuity approximation for the case of Muller's 
ratchet}

We wish to compute the steady state proportions $f_k(t)$ from 
Eqs.~\eqref{eq:balance-near-edgeI} and~\eqref{eq:balance-near-edgeII}, under the 
periodic boundary condition
\begin{equation}\label{eq:periodic-boundary}
  f_{k-1}(0)=f_k[1/(Uv)]\,.
\end{equation}
We introduce the generating function
\begin{equation}
G(\lambda,t)=\sum_{n\ge0}f_{k_0+n}(t)\lambda^n.
\label{defG}
\end{equation}
From Eqs.~\eqref{eq:balance-near-edgeI} and~\eqref{eq:balance-near-edgeII}, we 
obtain
\begin{equation}
\frac{\partial}{\partial t} G(\lambda,t)=-S G(\lambda,t)+U\lambda G(\lambda,t),
\end{equation}
so that
\begin{equation}
G(\lambda,t)=G(\lambda,0)e^{(U\lambda-S) t}.
\label{G1}
\end{equation}
We find $G(\lambda,0)$ from Eq.~\eqref{eq:periodic-boundary} and 
Eq.~\eqref{eq:balance-near-edgeI}
\begin{align}
\nonumber
  G(\lambda,0)&={1\over\lambda}G\Big(\lambda,{1\over Uv}\Big)
               -{1\over\lambda}f_{k_0}\Big({1\over Uv}\Big)\\
              &={1\over\lambda}e^{{U\lambda-S\over Uv}} G(\lambda,0)-
{1\over\lambda}
                    f_{k_0}(0)e^{-{S\over Uv}}.
\end{align}
Finally, using the fact that $S=Uv\ln(e/v)$, we find
\begin{equation}
G(\lambda,0)={f_{k_0}(0)\over e^{\lambda\over v}-\lambda e^{S\over Uv}}
={f_{k_0}(0)\over e^{\lambda\over v}-{\lambda e\over v}}.
\label{G2}
\end{equation}
The generating function $G(\lambda,t)$ contains all the information about the 
$f_k(t)$. By expanding Eqs.~\eqref{G1} and~\eqref{G2} in powers of $\lambda$ and 
identifying the terms with the coefficients in Eq.~\eqref{defG}, we recover all 
the functions $f_{k_0+n}(t)$:
\begin{align}
f_{k_0}(t)&=f_{k_0}(0)e^{-St},\qquad
f_{k_0+1}(t)=f_{k_0}(0){Uvt+e-1\over v}e^{-St},\\
f_{k_0+2}(t)&=f_{k_0}(0){(Uvt)^2/2+Uvt(e-1)+e^2-2e+1/2\over v^2}e^{-St},
\end{align}
etc. We obtain the large~$n$ behaviour of $f_{k_0+n}$ from the singularities of 
$G(\lambda,t)$: The first singularity is for $\lambda=v$, so that $f_{k_0+n}(t)$ 
must increase like $1/v^n$ for large~$n$. More precisely, to obtain the asymptotic 
behavior of $f_{k_0+n}(t)$, we study the divergence of $G(\lambda,t)$ around 
$\lambda\approx v$. We find
\begin{equation}
G(\lambda,t)=f_{k_0}(0){e^{(U\lambda-S) t}\over e^{\lambda\over
v}-e\times{\lambda
\over v}}=f_{k_0}(0)e^{(Uv-S) t}{2\over e} \Big({1\over(1-\lambda/v)^2}
+ 
{(1/3)-U v t \over 1 - \lambda/v}+\psi(\lambda,t)\Big),
\end{equation}
where the remaining part $\psi(\lambda,t)$ has no singularity at
$v=\lambda$. Using $\sum_{n\ge0} x^n=1/(1-x)$ and $\sum_{n\ge0} (n+1)x^n=1/(1-
x)^2$, we find
\begin{equation}
f_{k_0+n}(t)=f_{k_0}(0)e^{(Uv-S) t}{2\over e} \Big({n+1\over
v^n}+{(1/3)-Uvt\over v^n} +\delta_n(t)\Big),
\end{equation}
where $\delta_n(t)$ is defined by
$\psi(\lambda,t)=\sum_{n\ge0}\delta_n(t)\lambda^n$.  Except for $\lambda=v$, where 
$\psi(\lambda,t)$ is finite, the singularities of $\psi(\lambda,t)$ are the same 
as the singularities of $G(\lambda,t)$. We checked numerically that the next 
singularity is at $\lambda=8.07557e^{\pm i 1.1783}v$. Therefore, $\delta_n$ must 
decay faster than $1/(8v)^n$ for large $n$. Finally, for large $n$,
\begin{equation}
f_{k_0+n}(t)=f_{k_0}(t)e^{U v t}{2\over e v^n} \Big(n+{4\over3}-Uvt+{\cal
O}(1/8^n) \Big).
\end{equation}

We can now compare the proportions $f_k$ at, for instance, time~$t=1/(2Uv)$ (that 
is, in the middle of the cycle),
\begin{equation}\label{eq:comparison1}
\ln f_{k_0+n}\Big({1\over2Uv}\Big)-\ln f_{k_0}\Big({1\over2Uv}\Big)=
-n\ln v+\ln\Big[{2\over\sqrt{e}}\Big(n+{5\over6}\Big)\Big]+{\cal
O}(1/8^n),
\end{equation}
or, alternatively, compute the average of $\ln f_k$ over one cycle:
\begin{equation}\label{eq:comparison2}
  \overline{\ln f_{k_0+n}(t)}-\overline{\ln f_{k_0}(t)}
         =-n\ln v+\ln\Big[{2\over\sqrt{e}}\Big(n+{4\over3}\Big)\Big]+
             \Big(n+{1\over3}\Big)\ln\Big[1+{1\over n+{1\over3}}\Big]-1
               +{\cal O}(1/8^n).
\end{equation}
Both these asymptotic expressions give surprisingly good results already
for $n\ge1$. To order $1/n$, the right hand sides of Eqs.~\eqref{eq:comparison1} 
and~\eqref{eq:comparison2} are identical, and are given by
\begin{equation}
  -n \ln v + \ln n + \ln\Big(\frac{2}{\sqrt{e}}\Big)+\frac{5}{6n} + {\cal 
O}(1/n^2)\,.
\end{equation}

\subsection{Correction to the continuity approximation for the case of adaptive 
evolution}

As explained in the main text, deleterious mutation can be neglected for large 
$V,\ V\gg U$. We start from Eq.~\eqref{eq:det-expanded-Ualpha}, which we write as
\begin{equation}\label{evol}
f_{k}(t+1)=f_k(t)+ U_b f_{k+1}(t)-S f_k(t)\qquad\text{for $k\ge k_0$,}
\end{equation}
where $S=U+s(k-\kav)$. For $k<k_0$, we have $f_k(t)=0$. Strictly speaking, at each 
time step $t+1$ a proportion $U_b f_{k_0}(t)$ of the population arrives at site 
$k_0-1$. Yet these individuals are immediately removed from the population by 
genetic drift, because their number is too small. However, as time goes on, 
$f_{k_0}(t)$ grows, and at some point the number of individuals reaching ${k_0-1}$ 
is large enough to survive genetic drift. A new site is occupied and the value of 
$k_0$ decreases.

Let $t=0$ is the time at which a new clone is established at $k_0$. The average 
time the traveling wave moves one notch in $k$ is $1/V$, where $V$ is the average 
substitution rate. We approximate the process of adding new best-fit clones by a 
periodic process, i.e., a new clone will be established at site $k_0-1$ after a 
time interval of length $1/V$, a clone at site $k_0-2$ after another interval 
$1/V$, and so on. Then we have
\begin{equation}\label{periodic}
f_k(1/V)=f_{k+1}(0)\qquad\text{for $k\ge k_0$.}
\end{equation}
From the reasoning about the stochastic cutoff in the case of adaptive evolution, 
we know that a new clone starts to grow deterministically once it exceeds the 
characterstic threshold $f_k(t)\sim 1/(|S|N)$. Furthermore, we know from Eq.~\eqref{eq:def-fmax} that the adjacent class 
at this point in time is of size $f_{k+1}(t)\approx 1/(2NU_b)$. Thus we write
\begin{equation}\label{slope}
f_{k_0}(0)=\frac{C}{|S|N},\qquad f_{k_0}(1/V)=f_{k_0+1}(0)=\frac{1}{2NU_b}\,.
\end{equation}
where $C\sim 1$ is an undetermined numeric constant that, as we show below, does 
not affect much the final result in the parameter range we study.

By analogy with the derivation of the wave equation, we replace the difference 
equation \eqref{evol} by a differential equation in $t$. However, we do not make 
the continuity approximation in $k$. As in the case of the continuity correction 
for Muller's ratchet, for $k$ close the edge of distribution, we assume that $S$ 
is a constant independent of $k$, $S=U+sx_0$, where $x_0$ is given by 
Eq.~\eqref{eq:x-zero-AE}. Thus, we use
\begin{equation}\label{evol2}
{\partial f _{k}(t)\over\partial t}= U_b f_{k+1}(t)-S f_k(t)\qquad\text{for $k\ge 
k_0$}\,.
\end{equation}

We look for solutions of the form
\begin{equation}
f_{k_0+n}(t)=\sum_\lambda a_\lambda(t)e^{-\lambda n}\,.
\end{equation}
From Eq.~\eqref{evol2}, we obtain
\begin{equation}
a_\lambda(t) = a_\lambda(0)e^{(U_be^{-\lambda}-S)t}\,.
\end{equation}
From the periodicity condition, Eq.~\eqref{periodic}, we further obtain
\begin{equation}
a_\lambda(0)e^{(U_be^{-\lambda}-S)/V}=a_\lambda(0)e^{-\lambda}\,,
\end{equation}
so that, for each $\lambda$, either $ a_\lambda(0)=0$, or
\begin{equation}\label{eq:lambda-condition}
 S=U_be^{-\lambda}+\lambda V\,.
\end{equation}
Therefore,
\begin{equation}\label{generic1}
f_{k_0+n}(t)=\sum_{\lambda}
a_{\lambda}(0)e^{-\lambda(n+Vt)}\,.
\end{equation}
where the sum is over $\lambda$ that satisfy Eq.~\eqref{eq:lambda-condition}.

To make any further progress, we have to find the values of $\lambda$. The 
function $U_be^{-\lambda}+\lambda V$ has the only minimum
$-V[\ln(V/U_b)-1]$, which is reached for $\lambda=-\ln(V/U_b)$. From 
Eq.~\eqref{eq:x-zero-AE} and the definition of $S$, we see that $S=-V[\ln(V/U_b)-
1]+U$. Since, in the adaptation limit, we assume that $V\gg U$, 
Eq.~\eqref{eq:lambda-condition} will be satisfied for two values of $\lambda$ that 
are slightly larger and slightly smaller than $-\ln(V/U_b)$, as given by
\begin{align}
  \lambda^+ &= -\ln(V/U_b) + \sqrt{2U/V}\,,\\
  \lambda^- &= -\ln(V/U_b) - \sqrt{2U/V}\,.
\end{align}
If we now insert the expressions for $\lambda^+$ and $\lambda^-$ into 
Eq.~\eqref{generic1} and expand to first power in $\sqrt{U_b/V}$, we find
\begin{equation}\label{generic2}
f_{k_0+n}(t)= \Big[A+B(n+Vt)\Big]e^{\ln(V/U_b)(n+Vt)}\,,
\end{equation}
where we have introduced $A=a_{\lambda^+}(0)+a_{\lambda^-}(0)$ and 
$B=[a_{\lambda^-}(0)-a_{\lambda^+}(0)]\sqrt{2U/V}$. Note that we will not need 
these specific expressions for $A$ and $B$. Instead, we derive $A$ and $B$ 
directly from Eq.~\eqref{slope} and obtain
\begin{equation}
A=\frac{C}{|S|N},\qquad B=\frac{1}{N}\frac{|S|-2CV}{2|S|V}\,.
\end{equation}

Putting everything together, we can obtain the $f_k(t)$ at mid-period
[$t=(2V)^{-1}$], as given by
\begin{equation}
\ln f_{k_0+n}\left(1\over2V\right)-\ln f_{k_0}\left(1\over2V\right)
=n\ln(V/U_b)+\ln\left[1+2n-\frac{8CV}{|S|+2CV}n\right].
\end{equation}
If $|S|\gg V$, which is the case when $\ln(V/\Ub)\gg 1$, we can write the second 
term on the right-hand side as
\begin{equation}\label{eq:discont-corr-adaptation}
\ln\left[2n+1+{\cal O}\Big(\frac{1}{\ln(V/\Ub)}\Big)\right]\,.
\end{equation}

\end{appendix}
\cleardoublepage

\begin{table}
\begin{center}
\begin{tabbing}
Symbol \=~~~\= Definition \\[-.2cm]
\rule{5.5in}{.5pt}\\
\>$\alpha_k$\'\> ratio of the beneficial mutation rate to the total mutation rate 
in\\
 \>\>class $k$; ~~$\alpha_k=\mub{}k/U$ \\
\>$\alpha$\'\> effective ratio of the beneficial mutation rate to the total 
mutation\\
        \>\> ~~ rate; $\alpha=\mub{}\kav/U$\\
\>$f_k(t)$\'\> frequency of a class of genomes with mutation number $k$ at time 
$t$ \\
\>$\phi$\' \> logarithm of genome frequency; $\phi=\ln f$ \\
\>$k$\'\> the number of less-fit alleles in a genome, as compared to the best\\
        \>\> ~~possible genome \\
\>$k_0$\'\> minimum value of $k$ in the population \\
\>$\kav$\'\> effective mutational load generating a fitness equivalent to\\
        \>\> ~~ the mean fitness of the population, $\kav=-\frac{1}{s}\ln(\sum_k 
e^{-ks}f_k)$\\
\>$L$\'\> length of genome (number of sites) \\
\>$\mu$\'\> mutation rate per site \\
\>$N$\'\> haploid population size (number of genomes) \\
\>$\rho(f, t)$\'\> probability that a class of genomes has frequency $f$ at time 
$t$ \\
\>$s$\'\> selection coefficient; relative fitness gain/loss per mutation \\
\>$S$\'\> effective coefficient of selection against the best-fit class in a 
population;\\
  \>\> ~~ $S=U+s(k_0-\kav)$\\
\>$\sigma$\'\> rescaled selection coefficient; $\sigma=s/U$ \\
\>$t$\'\> time (in generations) \\
\>$U$\'\> effective mutation rate per genome per generation; $U=\mu L$ \\
\>$\Ub$\'\> beneficial mutation rate per genome per generation; $\Ub=(k/L)U$\\
\>$u$\'\> $u=e^{\phi'(x_0)}$ \\
\>$V$\'\> average substitution rate of beneficial mutations; $V=-d\kav(t)/dt$ \\
\>$v$\'\> normalized ratchet rate (substitution rate of deleterious mutations);\\
        \>\> ~~$v= (1/U)d\kav(t)/dt$\\
\>$\Vark$\'\> variance of $k$ in the population\\
\>$x$\'\> relative mutational load of a class; $x=k-\kav$ \\
\>$x_0$\'\> minimum value of $x$ for highest-fitness class \\[-.2cm]
\rule{5.5in}{.5pt}\\
\end{tabbing}
\end{center}
\caption{Variables used in this work.}\label{tab:variables}
\end{table}

\cleardoublepage

\end{document}